\documentclass[11pt]{article}  

\usepackage{amsmath, amsthm, amssymb}
\usepackage{graphicx}
\usepackage{bm} 

\usepackage[margin=1.2in]{geometry}

\usepackage{soul}
\usepackage{times}

\usepackage{pdfpages}

\theoremstyle{definition}

\theoremstyle{remark}

\newcommand{\var}{\text{var}}

\newcommand{\ee}{e}

\newcommand{\vvec}[1]{{\bm #1}}

\usepackage{color}
\definecolor{gray}{gray}{0.5}

\title{\vspace{-.8in}Evaluating the extent to which homeostatic plasticity learns to compute prediction errors in unstructured neuronal networks}

\author{Vicky Zhu and Robert Rosenbaum\footnote{This work was supported by the Air Force Office of Scientific Research (AFOSR) award number FA9550-21-1-0223 and NSF awards DMS-1517828  and DMS-1654268.}}

\begin{document}
\maketitle

\begin{abstract}

The brain is believed to operate in part by making predictions about sensory stimuli and encoding deviations from these predictions in the activity of ``prediction error neurons.''  This principle defines the widely influential theory of predictive coding. The precise circuitry and plasticity mechanisms through which animals learn to compute and update their predictions are unknown. Homeostatic inhibitory synaptic plasticity is a promising mechanism for training neuronal networks to perform predictive coding. Homeostatic plasticity causes neurons to maintain a steady, baseline firing rate in response to inputs that closely match the inputs on which a network was trained, but firing rates can deviate away from this baseline in response to stimuli that are mismatched from training. We combine computer simulations and mathematical analysis systematically to test the extent to which randomly connected, unstructured networks compute prediction errors after training with homeostatic inhibitory synaptic plasticity. We find that homeostatic plasticity alone is sufficient for computing prediction errors for trivial time-constant stimuli, but not for more realistic time-varying stimuli. We use a mean-field theory of plastic networks to explain our findings and characterize the assumptions under which they apply.

\end{abstract}

\section{Introduction}

Cortical neuronal networks can make predictions about sensory stimuli and detect errors about these predictions. For example, in the visuomotor system, head movements produce predictable flows of an animal's visual scene. Visual cortical circuits learn predictable associations between bottom-up input from the visual stream and top-down input from the motor system. Violations of the learned predictions, known as ``mismatched stimuli'' or ``prediction errors'', produce distinct responses in visual cortical neurons, which can help the animal distinguish between self-driven and externally driven movements of its visual scene~\cite{keller2012sensorimotor,leinweber2017sensorimotor,attinger2017visuomotor}. 

The idea that the brain uses predictions and prediction errors to encode and interpret sensory information dates back to 19th century work by Helmholz~\cite{von1867handbuch,keller2018predictive} and underlies more general theories of neural function such as predictive coding, predictive processing, active inference, and the free energy principle~\cite{rao1999predictive,friston2010free,clark2015surfing,keller2018predictive}.  
The question of how neural circuits compute prediction errors and how they learn predictions through biologically plausible synaptic plasticity rules is not  settled, but some theories have been put forward~\cite{wacongne2012neuronal,bastos2012canonical,rao200216,bogacz2017tutorial,whittington2019theories,hertag2020learning,schulz2021generation}.

Cortical neurons are highly interconnected, even within a single cortical area and layer. This dense, recurrent, and intralaminar connectivity shapes the intrinsic dynamics and stimulus responses of local cortical circuits. The nonlinear firing rate dynamics that arise from this recurrent connectivity can interact with the slower dynamics of synaptic plasticity in complex ways. Homeostatic inhibitory synaptic plasticity is a widely observed and widely studied type of synaptic plasticity~\cite{castillo2011long,Vogels2011,luz2012balancing,Vogels2013,Hennequin2017,schulz2021generation,capogna2021ins} in which the strength of inhibitory synapses are adjusted in an activity-dependent manner that tends to push the postsynaptic neurons' firing rates toward a homeostatic baseline targets. Simulations and theoretical analyses of mathematical models of homeostatic inhibitory plasticity show that, while firing rates are near their targets in response to stimuli on which the network has been trained, firing rates deviate from their targets in response to unfamiliar stimuli in these models~\cite{Vogels2011,baker2020nonlinear,hertag2020learning,hertag2021prediction,schulz2021generation,akil2021balanced}.

As in related computational work~\cite{hertag2020learning,hertag2021prediction,schulz2021generation}, we conjectured that homeostatic inhibitory plasticity could learn to perform some type of predictive coding. In particular, if the external input to a neural population were formed from bottom-up and top-down stimuli, then homeostatic plasticity in the network would naturally learn to produce baseline activity in response to ``matched''  top-down and bottom-up pairings ({\it i.e.}, pairings that are similar to those on which the network was trained). On the other hand, ``mismatched'' pairings ({\it i.e.}, pairings from outside the training distribution) would produce firing rate responses that are further from the homeostatic baseline. In this sense, the  network should learn to encode prediction errors ({\it i.e.}, errors in the ability to predict top-down input from bottom-up input or vice versa) in the deviation of the firing rates from their baseline. Importantly, and in contrast to previous work~\cite{hertag2020learning,hertag2021prediction,schulz2021generation}, we conjectured that the network should not need to be imparted with any special structure or architecture to learn this computation since homeostatic plasticity should naturally achieve this result due to its tendency to produce baseline responses to stimuli on which the network was trained, but not in response to novel stimuli. 


To test our conjecture, we used an unstructured, recurrent, spiking neuronal network model endowed with a homeostatic inhibitory plasticity rule receiving two sources of external input, modeling top-down and bottom-up stimuli. We trained the network with given patterns of top-down and bottom-up  pairings, interpreted as ``matched'' stimuli, before presenting a ``mismatched'' stimulus that deviated from the pairings used during training. Numerical simulations showed that the network reliably produced baseline firing rates for a fixed pair of bottom-up and top-down inputs during training, and deviated from baseline in response to a mismatched stimulus. A mean-field, firing rate model and a mathematical analysis using a separation of timescales helped reveal the dynamics underlying these numerical simulations. Hence, homeostatic plasticity learned to compute prediction errors whenever top-down and bottom-up stimuli are fixed during training. However, useful predictive coding algorithms should learn to detect relationships between time-varying top-down and bottom-up inputs. We generalized our input model to vary the intensity of top-down and bottom-up inputs in unison. An effective learning algorithm should learn to detect a prediction error whenever the intensity changes out of unison.  To our surprise, our spiking network with homeostatic synaptic plasticity was unable to learn to detect this type of prediction error, even in a relatively simple (time-varying) setting. Going back to our mean-field analysis helped to clarify how and why the model failed to perform predictive coding in this setting after succeeding in the simpler (time-constant) setting. 

We conclude that homeostatic inhibitory synaptic plasticity alone is not sufficient to learn and perform non-trivial predictive coding in unstructured neuronal network models. Previous theoretical work shows that network models that carefully account for the connectivity structure of multiple inhibitory subtypes are able to learn prediction errors using homeostatic plasticity, even for inputs where top-down and bottom-up input co-vary in intensity~\cite{hertag2020learning,hertag2021prediction}. Hence, the failure of our model in this scenario implies that network structure is critical for successfully learning predictive coding tasks with homeostatic plasticity.

\section{Results}

\subsection{Spiking network model description}

We consider a computational model of a local cortical circuit composed of $N=5000$ randomly connected exponential integrate-and-fire (EIF) spiking neuron models ($N_e=4000$ of which are excitatory and $N_i=1000$ inhibitory)~\cite{brette2005adaptive,gerstner2014neuronal}. 
The membrane potentials of neuron $j$ in population $a=e,i$ obeys
\begin{equation}\label{E:dV}
\tau_m \frac{d V^a_j}{dt}=-(V^a_j-E_L)+D_Te^{(V^a_j-V_T)/D_T}+I^a_j(t)
\end{equation}
with the added condition that each time $V_k(t)$ crosses a threshold at $V_{th}$, it is reset to $V_{re}$ and a spike is recorded. 
The synaptic input to neuron $j$ in population $a$ is modeled by 
\[
I^a_j(t)=X^a_j(t)+\sum_{b=e,i}\sum_{k=1}^N J^{ab}_{jk} \alpha_b(t-t^b_{n,k})
\]
where $X^a_j(t)$ models external synaptic input, $J^{ab}_{jk}$ is a synaptic weight, $t^b_{n,k}$ is the time of the $n$th spike of neuron $k$ in population $b$, and $\alpha_b(t)=(1/\tau_b)e^{-t/\tau_b}H(t)$ is a synaptic filter with $H(t)$  the Heaviside step function. 

Initial connectivity in the model is random (connection probability $p=0.1$) with initial weights, $J_{jk}^{ab}$, determined only by pre- and post-synaptic neuron type ($J_{jk}^{ab}=j_{ab}$ for connected neurons). Excitatory connectivity, $J_{jk}^{ae}$, remained fixed, but inhibitory connectivity evolves according to a homeostatic, inhibitory spike-timing-dependent plasticity (iSTDP) rule~\cite{Vogels2011,Vogels2013,Hennequin2017,akil2021balanced}. Specifically, each time that neuron $j$ in population $a=e,i$ spikes (which occurs at times $t^a_{n,j}$), the inhibitory synaptic weights targeting that neuron are updated according to 
\[
J^{ai}_{jk}=J^{ai}_{jk}-\eta_a x^i_k(t_{j,n}^a)
\]
where $\eta_a$ is a learning rate and recall that $t_{j,n}^a$ is the time of the $n$th spike of neuron $j$ in population $a=e,i$. 
Additionally, each time inhibitory neuron $k$ spikes, its outgoing synaptic weights are updated according to
\[
J^{ai}_{jk}=J^{ai}_{jk}-\eta_a \left(x^a_j(t_{k,n}^i)-2r_0^a\right)
\]
where $t_{k,n}^i$ is the time of the $n$th spike of inhibitory neuron $k$. 
The time series, $x^a_j(t)$ are defined by the differential equation
\[
\tau_{STDP}\frac{dx^a_j}{dt}=-x^a_j
\]
in addition to the rule that $x^a_j(t)$ is incremented each time that neuron $j$ in population $a=e,i$ spikes according to,
\begin{equation}\label{E:dx}
dx^a_j(t^a_{j,n})\gets dx^a_j(t^a_{j,n})+\frac{1}{\tau_{STDP}}.
\end{equation}
As a result, $x^a_j(t)$ estimates the firing rate of neuron $j$ in population $a$ by performing an exponentially-weighted sliding average of the spike density. 
This plasticity rule tends to push excitatory and inhibitory firing rates toward their target rates, $r_0^e$ and $r_0^i$, respectively (see \cite{Vogels2011,Vogels2013,Hennequin2017,baker2020nonlinear,akil2021balanced} and the mean-field theory presented below).

We are interested in understanding the extent to which  such networks can learn to perform predictive coding~\cite{rao1999predictive,bogacz2017tutorial,keller2018predictive}. 
More specifically, we reasoned that neurons would spike close to their target rates in response to stimulus patterns similar to those on which they were trained, but deviate from the target rates in response to stimuli that deviate from the from the training stimuli. In other words, the deviation of firing rates from their targets should encode  a ``prediction error,'' {\it i.e.}, a deviation of the inputs from the patterns that appeared during training. 


\subsection{Prediction errors after training on time-constant inputs to multiple sub-populations}

 \begin{figure*}
 \centering{
 \includegraphics[width=6in]{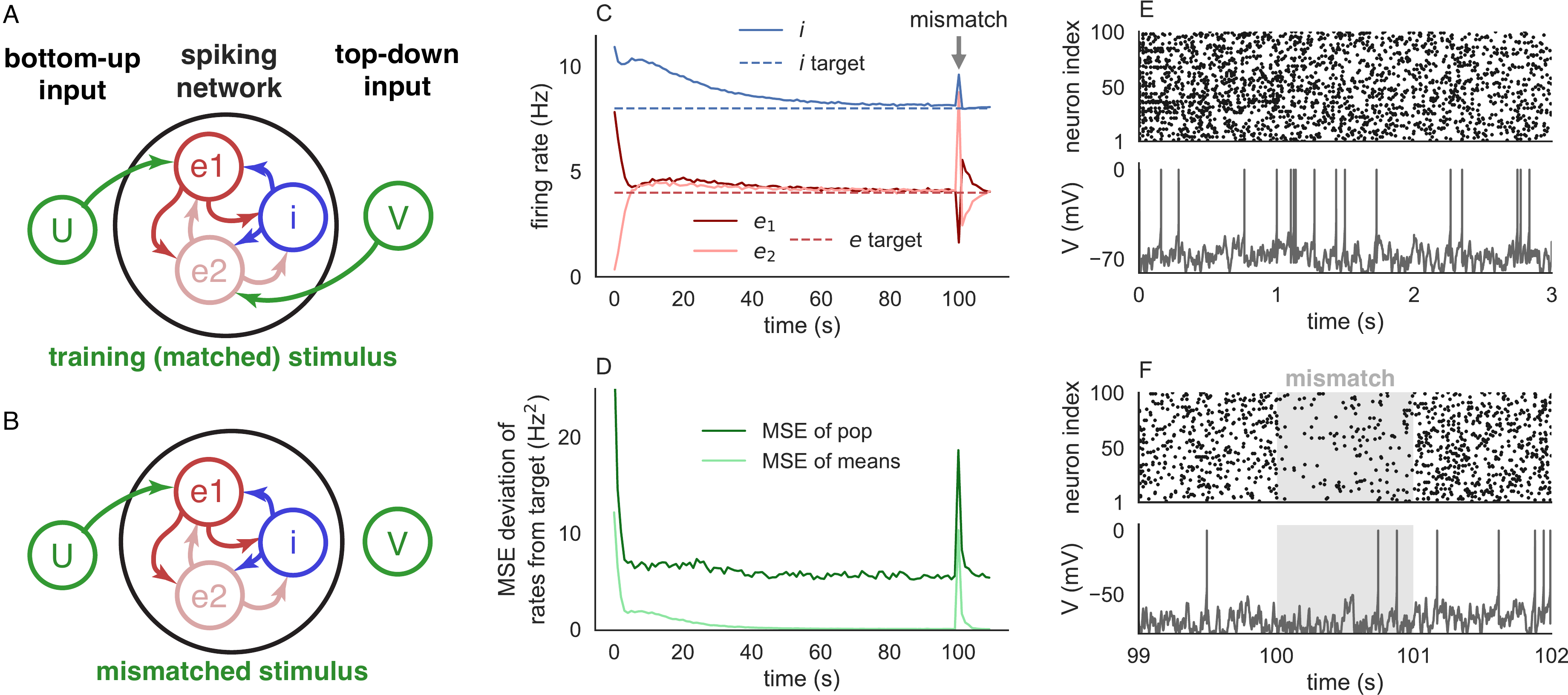}
 }
 \caption{{\bf Prediction errors after training on time-constant inputs to multiple sub-populations.} {\bf A,B)} Network diagram with ``training'' and ``mismatch'' stimuli respectively. A randomly connected, recurrent spiking neural network of $N=5000$ neurons consisted of two excitatory sub-populations ($e_1$ and $e_2$) and one inhibitory ($i$) population. During the first 100s of the simulation, the network received a ``training'' stimulus in which $e_1$ and $e_2$ received extra external input modeling bottom-up and top-down stimuli respectively (A). Then a ``mismatch'' stimulus was introduced for 1s by removing the top-down stimulus to population $e_2$.  {\bf C)} Homeostatic inhibitory synaptic plasticity caused population-averaged firing rates to converge to their targets during training, but they deviated from their targets in response to the mismatch stimulus. {\bf D)} The deviation of the mean firing rates from their targets ($MSE_{mean}$) and the mean deviation of individual neurons' firing rates ($MSE_{pop}$) quantify the deviation of firing rates from their targets. {\bf E,F)} Raster plots (top) and membrane potential (bottom) of a random subset of neurons from population $e_1$.     }
 \label{Fig1}
 \end{figure*}

For illustrative purposes, we first considered a simple input model for which the excitatory population was divided into two sub-populations, $\ee_1$ and $\ee_2$, with $N_{\ee_1}=N_{\ee_2}=2000$ neurons in each sub-popuation (Figure~\ref{Fig1}A,B). Recurrent connectivity did not depend on sub-population membership, so the network was completely unstructured.  During training, each neuron in populations $\ee_1$ and $\ee_2$ received external stimuli of the form (Figure~\ref{Fig1}A)
\begin{equation}\label{E:Xmatched}
\left.
\begin{aligned}
X_{\ee_1}&=X_e^0+U\\
X_{\ee_2}&=X_e^0+V
\end{aligned}
\,\right\rbrace\textrm{ matched}
\end{equation}
where $X_e^0$ is a baseline input that assures neurons spike at reasonable rates, $U$ is a perturbation modeling bottom-up input, and $V$ is a perturbation modeling top-down input. We used positive bottom-up input and negative top-down input, 
\begin{equation}\label{E:UVhomog}
\begin{aligned}
U&=X_e^0/5\\
V&=-X_e^0/5,
\end{aligned}
\end{equation}
but our results are not sensitive to this specific choice of inputs. 
We refer to this as a ``matched'' stimulus because it defines the matching of bottom-up with top-down stimuli that the network is trained on. 
After training on matched stimuli, we  modeled mismatched stimuli  by the absence of top-down input (Figure~\ref{Fig1}B),
\begin{equation}\label{E:Xmismatched}
\left.
\begin{aligned}
X_{\ee_1}&=X_e^0+U\\
X_{\ee_2}&=X_e^0
\end{aligned}
\,\right\rbrace\textrm{ mismatched}.
\end{equation}
We refer to these stimuli as ``mismatched'' because the top-down and bottom-up inputs are mismatched when compared to the ``matched'' pairings used to train the network. Mismatched stimuli could also be modeled by an absence of bottom-up input, or any other deviation from the inputs used for training.


We hypothesized that, after training on matched stimuli, the network would  produce firing rates close to the target rates in response to matched stimuli and produce firing rates further from the target rates in response to mismatched stimuli. 

%


At the beginning of the simulation mean excitatory and  inhibitory firing rates deviated from their targets, but inhibitory plasticity pushed them  toward their targets over the course of tens of seconds (Figure~\ref{Fig1}C). 
After 100s of training on matched stimuli, we tested a mismatched stimulus for 1s. Consistent with our hypothesis, mean firing rates of each population were further from their targets in response to the mismatched stimulus (Figure~\ref{Fig1}C).

We quantified the distance of the firing rates from their targets from spiking network simulations using two methods. For the first method, we computed the MSE of the population-averaged firing rates (Figure~\ref{Fig1}D, light green),
\[
MSE_{mean}=\sum_{a=e_1,e_2,i}q_a(r_a-r^0_a)^2
\]
where $r^0_a$ is the target rate and $r_a$ the mean firing rate of each population averaged over neurons in that population and averaged over time windows of size $T=1$s. The coefficients $q_a=N_a/N$ represent the proportion of the network contained in each population ($q_{e_1}=q_{e_2}=0.4$ and $q_i=0.2$ for our network). Hence,  $MSE_{mean}$ weights the errors of larger sub-populations more heavily.

The $MSE_{mean}$ measures how far the population-average rates differ from their target rates, but does not measure the deviation of individual neurons' firing rates. Despite the fact that external input was constant across time and the simulations were deterministic (with the exception of ``quenched'' randomness from the random connectivity), neurons exhibited substantial variability in their spike timing and membrane potential dynamics (Figure~\ref{Fig1}E,F). These dynamics are characteristic of an asynchronous-irregular state~\cite{van2005course,vanVreeswijk:1996us,Amit:1997uj,vanVreeswijk:1998uz,Brunel:1999ua,Brunel:2000th,Renart2010}. 

To account for the deviation of individual neurons' firing rates from spike-timing variability in spiking network simulations, we also computed the MSE across the entire network (Figure~\ref{Fig1}D, dark green),
\[
MSE_{pop}=\frac{1}{N}\sum_{j=1}^N (r_j-r^0_j)^2
\]
where $r_j$ is the firing rate of neuron $j=1,\ldots,N$ and $r^0_j$ is its target rate. Both measures of MSE show a decrease during training and a sharp increase in response to the mismatched stimulus, but $MSE_{pop}$ is larger overall due to the spike-timing variability of each neuron. 

The results from the spiking network can be understood using a simpler dynamical mean-field model in which mean firing rates of each population are approximated by a system of differential equations, 
\begin{equation}\label{E:drdt}
 \tau\odot\frac{d{\bf r}}{dt}=-{\bf r}+f\left(W{\bf r}+{\bf X}\right)
\end{equation}
where $\tau=[\tau_{e_1}\;\; \tau_{e_2}\;\; \tau_i]^T$ is a vector of time constants, $\odot$ represents element-wise multiplication, and ${\bf r}=[r_{e_1}\;\; r_{e_1}\;\; r_i]^T$ is a vector approximating the mean firing rates of the two excitatory sub-populations and the inhibitory population.  Mean external input to each population is given by the vector
\[
{\bf X}=\left[\begin{array}{c}X_{e_1}\\X_{e_2}\\X_i\end{array}\right]
\]
and the recurrent connectivity matrix is defined by
\[
W=\left[\begin{array}{ccc}w_{e_1 e_1} & w_{e_1 e_2} & w_{e_1 i}\\ w_{e_2 e_1} & w_{e_2 e_2} & w_{e_2 i}\\w_{i e_1} & w_{i e_2} & w_{i i}\end{array}\right]
\]
where~\cite{pyle2016highly,pyle2017spatiotemporal,ebsch2018imbalanced,baker2019correlated,baker2020nonlinear,akil2021balanced}
\[
w_{ab}=N_bp_{ab}j_{ab}
\]
Here, $N_b$ is the number of neurons in population $b=e_1,e_2,i$ (so $N_{e_1}=N_{e_2}=N_e/2=2000$ and $N_i=1000$), $p_{ab}$ is the connection probability from population $b$ to population $a$, and $j_{ab}$ is the mean non-zero synaptic weight (mean of $J_{ab}^{jk}$ between connected neurons).  
The inhibitory entries, $w_{ai}$ for $a=e_1,e_2,i$, are negative and evolve according to
\begin{equation}\label{E:dwaidt}
\frac{dw_{ai}}{dt}=-\eta_a(r_a-r^a_0)r_i
\end{equation}
where $\eta_{a}$ sets the timescale of plasticity and $r_0^a$ is the target rate of population $a=e_1,e_2,i$.
For simplicity, we consider a rectified-linear f-I curve, 
\begin{equation}\label{E:fIcurve}
f(I)=\begin{cases}gI & I>0\\ 0 & I\le 0\end{cases}.
\end{equation}
The gain, $g$, was fit to spiking network simulations (see Materials and Methods).

 \begin{figure}
 \centering{
 \includegraphics[width=2.25in]{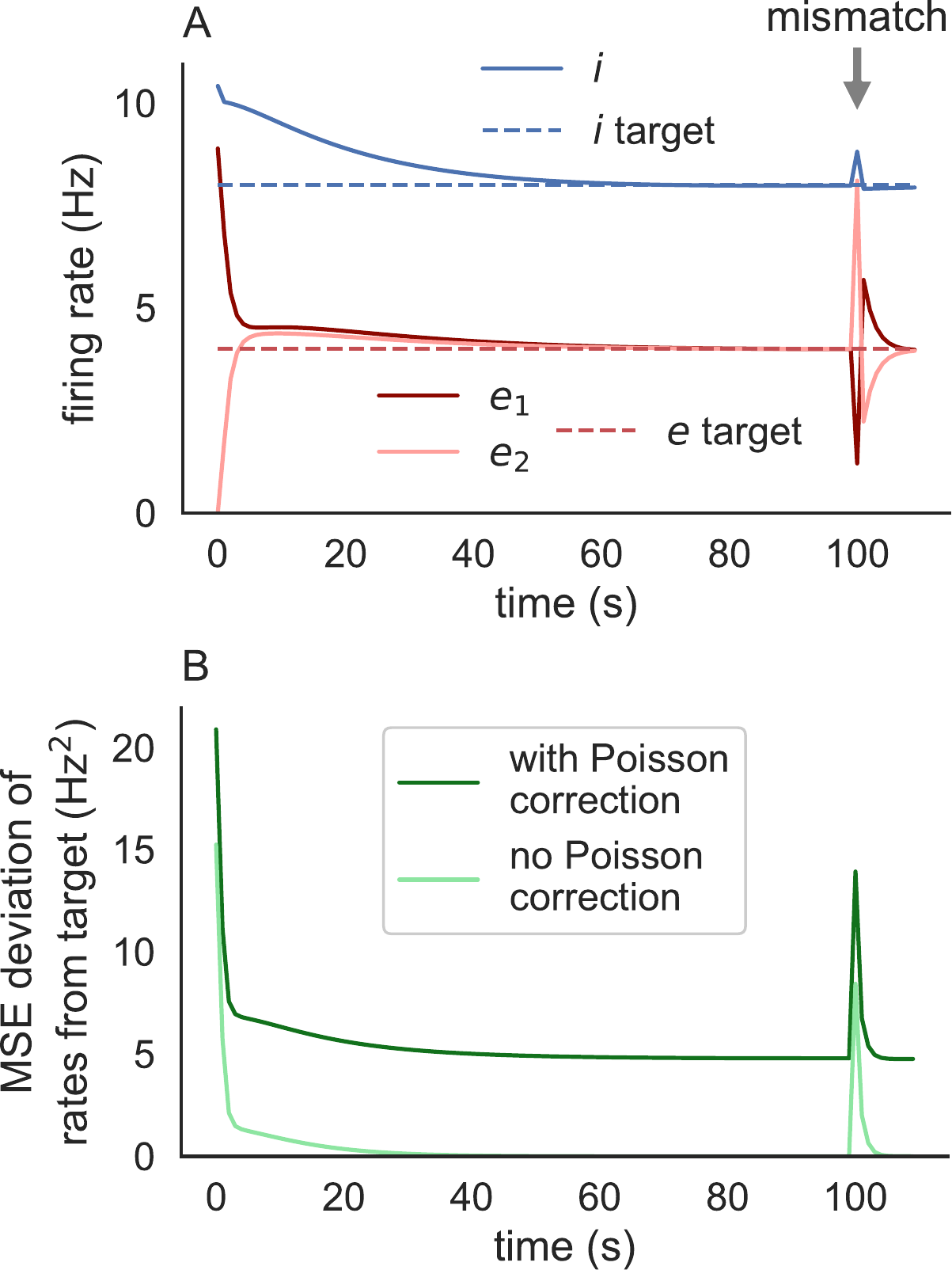}
 }
 \caption{{\bf A mean-field firing rate model captures the dynamics of the spiking network model.} {\bf A)} Firing rates of the mean-field firing rate model defined by Eqs.~\eqref{E:drdt} and \eqref{E:dwaidt}. Compare to Figure~\ref{Fig1}C. {\bf B)} MSE deviation of the firing rates from their targets ($MSE_{mf}$; light green) and the MSE with a Poisson correction ($MSE_{Poisson}$; dark green). Compare to Figure~\ref{Fig1}D.}
 \label{Fig2}
 \end{figure}

Simulating this model shows excellent agreement with the firing rates from the spiking network simulations (Figure~\ref{Fig2}) and the mean-field simulations are computationally more efficient than the spiking network simulations by a factor of 70 (6.0s for the mean-field simulation compared to $435.0$s for the spiking network simulation). 
The deviation of the firing rates in the mean-field rate model from their targets can be quantified by
\begin{equation}\label{E:MSEmf}
MSE_{mf}=\sum_{a=e_1,e_2,i}q_a (r_a-r^0_a)^2
\end{equation}
which is identical to $MSE_{mean}$ above except that $r_a$ represents the rate from the mean-field simulations instead of the mean firing rates from the spiking net simulations. Indeed, $MSE_{mf}$ closely matches $MSE_{mean}$ from the spiking network simulations (Figure~\ref{Fig2}B, compare to Figure~\ref{Fig1}C), demonstrating that the two models have similar mean-field dynamics. The value of $MSE_{pop}$ from the spiking network simulations does not have a direct analogue in the mean-field model, but under an assumption of Poisson-like spike-timing variability in the spiking network, $MSE_{pop}$ can be approximated by (see Materials and Methods for derivation)
\begin{equation}\label{E:MSEPoisson}
MSE_{Poisson}= MSE_{mf}+\frac{1}{T}\sum_a q_a r_a
\end{equation}
where $r_a$ is the firing rate of population $a=e_1,e_2,i$ from the mean-field model and $T$ is length of the time window over which firing rates are computed in the spiking network simulations. Specifically, $MSE_{Poisson}$ represents the population-level MSE ({\it i.e.}, $MSE_{pop}$) that would be produced by populations of Poisson spike trains with firing rates $r_a$. 
Indeed, $MSE_{Poisson}$ shows close agreement with $MSE_{pop}$ (Figure~\ref{Fig2}B, compare to Figure~\ref{Fig1}D), demonstrating that the deviation of $MSE_{pop}$ away from the values of $MSE_{mean}$ is consistent with Poisson-like spike-timing variability. 

This example shows that homeostatic inhibitory synaptic plasticity can train a network to detect mismatched stimuli, which is a form of predictive coding. To better understand how and why the network is able to detect mismatched stimuli, we consider a fixed point analysis via a separation of timescales. 

In the absence of plasticity ($W$ fixed, {\it e.g.}, $\eta_e=\eta_i=0$), fixed point firing rates would satisfy ${\bf r}_0=f(W{\bf r}_0+{\bf X})$. Taking the rectified linear f-I curve from the dynamical mean-field model, if there were a fixed point with positive rates ($r_a>0$ for all $a$) then it would be unique and given (as a function of $W$)  by
\begin{equation}\label{E:rqss}
{\bf r}(W)=\left[D-W\right]^{-1}\vvec X=A\vvec X
\end{equation}
where $D=(1/g)Id$ is a diagonal matrix, $I$ is the identity matrix, and $A=[D-W]^{-1}$. With $W$ fixed, the Jacobian matrix for the firing rate equation, Eq.~\eqref{E:drdt}, would be given by
\[
{\bf J}=g\left[\begin{array}{ccc} 
({w_{e_1e_1}-1})/{\tau_e} & {w_{e_1e_2}}/{\tau_e} & {w_{e_1i}}/{\tau_e}\\   
 {w_{e_1e_2}}/{\tau_e} & ({w_{e_1e_1}-1})/{\tau_e} & {w_{e_1i}}/{\tau_e}\\   
{w_{ie_1}}{\tau_i} & {w_{ie_2}}{\tau_i} &({w_{ii}-1})/{\tau_i}\end{array}\right]
\]
If the eigenvalues of this matrix have negative real part, then the fixed point given by Eq.~\eqref{E:rqss} is stable and globally attracting. 

\begin{figure*}
 \centering{
 \includegraphics[width=6in]{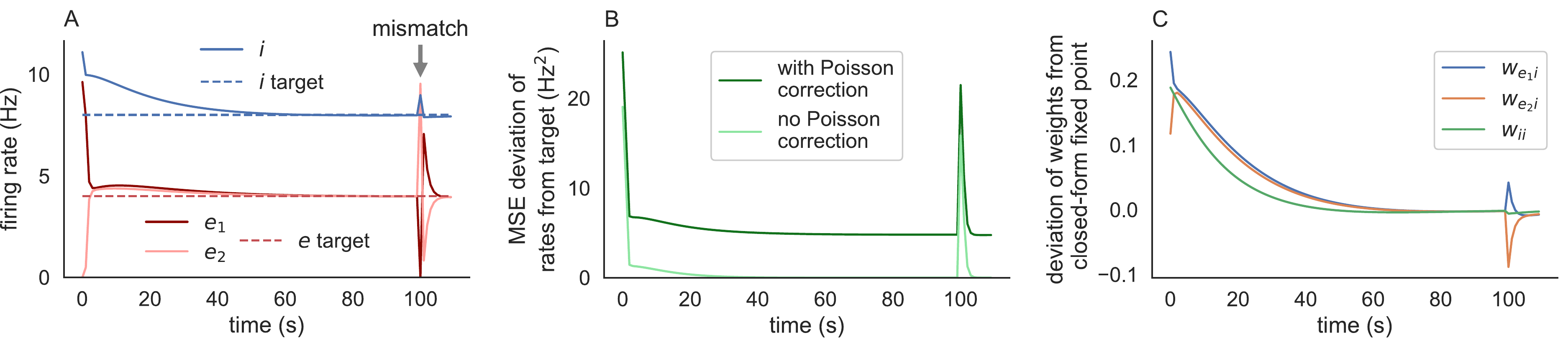}
 }
 \caption{{\bf Slow dynamics are captured by a separation-of-timescales approximation.} {\bf A)} Firing rates of the model defined by Eqs.~\eqref{E:dwrs}. Compare to Figures~\ref{Fig1}C and \ref{Fig2}A. {\bf B)} MSE deviation of the firing rates from their targets ($MSE_{mf}$; light green) and the MSE with a Poisson correction ($MSE_{Poisson}$; dark green) from the model defined by Eqs.~\eqref{E:dwrs}. Compare to Figures~\ref{Fig1}D and \ref{Fig2}B. {\bf C)} Deviation of the inhibitory weights, $w_{ai}$, from the fixed point values given in Eqs.~\eqref{E:wFP}.  }
 \label{Fig3}
 \end{figure*}

Due to plasticity, $W$ itself is time-dependent, so this fixed point analysis does not tell the full story. 
When plasticity is much slower than the firing rate dynamics ($\eta$ sufficiently small and $\tau$ sufficiently large, but $\eta$ should not be compared directly to $\tau$ because they have different dimensions), we can perform a separation of timescales under which $\vvec r$ relaxes to the quasi-steady-state value given by evaluating Eq.~\eqref{E:rqss} at the current value of $W$, while $W$ evolves more slowly according to Eq.~\eqref{E:dwaidt}. Putting this together, the separation of timescales approximation is defined by 
\begin{equation}\label{E:dwrs}
\begin{aligned}
\frac{dW}{dt}&=\left[\begin{array}{ccc} 0 &  0 & -\eta_e(r_{e_1}-r^e_0)r_i\\ 0 &  0 & -\eta_e(r_{e_2}-r^e_0)r_i\\ 0 & 0 & -\eta_i(r_i-r^i_0)r_i\end{array}\right]\\
\vvec r&=\left[\begin{array}{c}r_{e_1}\\ r_{e_2}\\ r_i\end{array}\right]=\left[D-W\right]^{-1}\vvec X=A\vvec X
\end{aligned}
\end{equation}
Note that this is a 3-dimensional dynamical system because $\vvec r$ is defined by a functional relationship instead of a differential equations. 
Solving Eqs.~\eqref{E:dwrs} directly gives similar results to the full mean-field model and is 482 times more computationally efficient than the full mean-field simulations (Figure~\ref{Fig3}A,B; $12.5\times 10^{-3}$s to simulate Eqs.~\eqref{E:dwrs} versus 6.0s for the full mean-field model) primarily because the slower dynamics allow for a larger time discretization (we used $dt=0.1$ms for the full mean-field and $dt=T=1$s to simulate Eqs.~\eqref{E:dwrs}). Simulating Eqs.~\eqref{E:dwrs} was 34751 times faster than the spiking network simulations. This speedup is not surprising given the lower dimension (2 versus 5000 dimensions) as well as the larger time discretization.

During training, $\vvec X$ is fixed to the ``matched'' value given by Eq.~\eqref{E:Xmatched}. During this phase, the slow-timescale system described by Eqs.~\eqref{E:dwrs} has a fixed point for which $\vvec r=\vvec r^0$ where 
\[
\vvec r^0=\left[\begin{array}{c}r_e^0\\ r_e^0\\ r_i^0\end{array}\right]
\] 
is a vector of the target rates from the plasticity rule. 
However, this expression gives the fixed point in terms of $\vvec r$ whereas the dynamical system is described by the dynamics of the entries of $W$. 
If the network converges to the target rates during training, then the weight matrix, $W$, for the slow system converges to a value, $W^0$ (or, equivalently,  $A$ converges to a value of $A^0$) that satisfies 
\begin{equation}\label{E:Wr0}
\left[D-W^0\right]^{-1}\vvec X^m=A^0\vvec X^m=\vvec r^0
\end{equation}
where  
\[
\vvec X^m=\left[\begin{array}{c}X_e^0+U\\X_e^0+V\\ X_i^0\end{array}\right]
\]
is the value of $\vvec X$ for matched stimuli. Eq.~\eqref{E:Wr0} is a system of three equations for three unknowns ($w_{e_1i}$, $w_{e_2i}$, $w_{ii}$) and its solution is given by
\begin{equation}\label{E:wFP}
\begin{aligned}
w_{e_1i}&=\frac{r_e^0-2 g r_e^0 w_{ee}-g (U+X_e^0)}{g r_i^0}\\
w_{e_2i}&=\frac{r_e^0-2 g r_e^0 w_{ee}-g (V+X_e^0)}{g r_i^0}\\
w_{ii}&=\frac{r_i^0-2 r_e^0 w_{ie}+X_i^0}{gr_i^0}
\end{aligned}
\end{equation}
Indeed, the weights converged toward these fixed point values during the training period (before the mismatch stimulus; Figure~\ref{Fig3}C). 


When the input is changed by a mismatched stimulus (so $\vvec X$ changes away from its value during training), firing rates deviate from their targets. Using the same quasi-steady state approximation, we can quantify the magnitude of this deviation as
\begin{equation}\label{E:dr}
\begin{aligned}
d\vvec r&:=\vvec r^{mm}-\vvec r^{0}\\
&=A^0\vvec X^{mm}-\vvec r^{0}\\
&=A^0(\vvec X^{mm}-\vvec X^m)\\
&=A^0d\vvec X
\end{aligned}
\end{equation}
where $\vvec r^{mm}$ is the vector of firing rates during a mismatched trial, $\vvec r^{0}=[r_e^0\;\; r_i^0]^T$ is the vector of target rates, and 
\[
d\vvec X=\vvec X^{mm}-\vvec X^m=\left[\begin{array}{c}0 \\ -V\\ 0\end{array}\right]
\]
is the perturbation of the external stimulus away from its training value during the mismatched trial. 
This derivation makes it clear that larger perturbations of the stimulus (larger values of $\|d\vvec X\|$) generally lead to larger deviations of the firing rates from their targets (larger values of $\|d\vvec r\|$). Here and elsewhere, $\|\cdot \|$ refers to the Euclidean norm.

Firing rate perturbations, $\|d\vvec r\|$, are especially large if the input perturbations, $d\vvec X$, point in a direction in which $A^0d\vvec X$ is large. Such directions correspond to the directions indicated by the largest eigenvalue(s) of $A^0$.  Since $A^0=\left[D-W^0\right]^{-1}$, when $W^0$ is much larger than $D$ in magnitude, these directions correspond to directions indicated by the smallest eigenvalue(s) of $W^0$. This phenomenon is an instance of ``imbalanced amplification'' in which a perturbation that points toward the nullspace or ``approximate nullspace'' of the connectivity matrix, $W^0$, is amplified by the network, see~\cite{ebsch2018imbalanced} for more in-depth explanations. 

Temporarily ignoring the direction of the perturbation, we can make the rough approximation that $\|d\vvec r\|$ is approximately proportional to $\|d\vvec X\|$. 
This rough approximation provides the intuition for mismatched responses shown in the simulations above. Put simply, mismatched responses are caused by the deviation of a stimulus away from its ``matched'' training value and the magnitude of the mismatched response increases with the magnitude of the input perturbation. 
While this intuition may seem trivial for this example, its extensions will help explain some non-trivial, counterintuitive results below. 

\subsection{Prediction errors after training on distributed, time-constant inputs}

The example above modeled a  stimulus that was homogeneous across each neural population, {\it i.e.}, every neuron in population $e_1$ received the same input and every neuron in population $e_2$ received the same input. Stimulus representations in cortical circuits can be distributed in an inhomogeneous way across neural populations~\cite{saxena2019towards}.

 \begin{figure*}
 \centering{
 \includegraphics[width=6in]{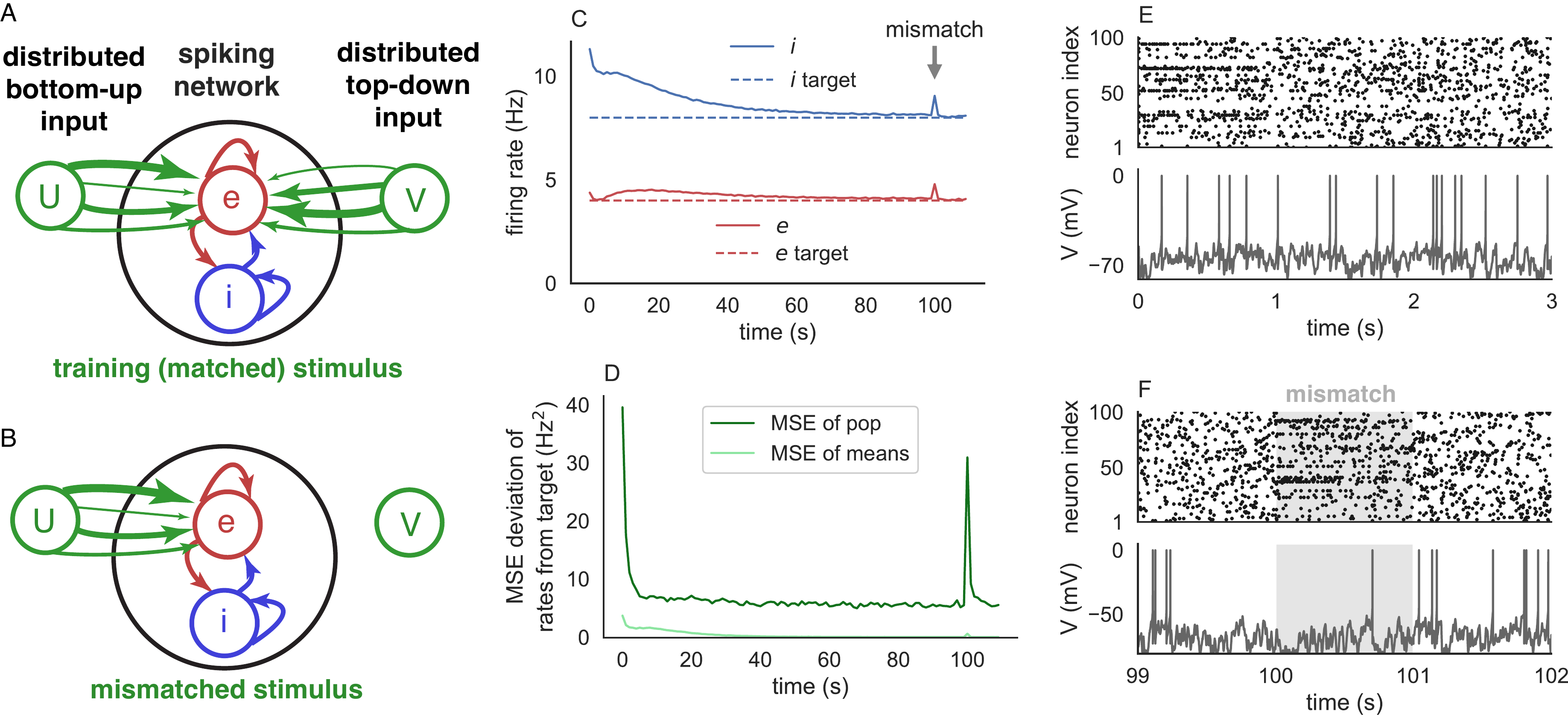}
 }
 \caption{{\bf Prediction errors after training on distributed time-constant inputs.} Same as Figure~\ref{Fig1} except bottom-up and top-down inputs were modeled as distributed stimuli using multivariate Gaussian inputs vectors (Eq.~\eqref{E:UVDist}).}
 \label{Fig4}
 \end{figure*}

We next considered a spiking network model with distributed bottom-up and top-down inputs (Figure~\ref{Fig4}A). 
As above, matched and mismatched stimuli were defined by the presence and absence of top-down input to population $e_2$ (Eqs.~\eqref{E:Xmatched} and \eqref{E:Xmismatched}) to match the bottom-up input to population $e_1$, but these inputs are heterogeneous vectors ($\vec U$ and $\vec V$) instead of homogeneous scalars ($U$ and $V$). Specifically, matched and mismatched stimuli to excitatory neurons were defined by
\begin{equation}\label{E:XDistmatched}
\left.
\begin{aligned}
X_{\ee}&=X_e^0+\vec U+\vec V\\
\end{aligned}
\,\right\rbrace\textrm{ matched}
\end{equation}
and
\begin{equation}\label{E:XDistmismatched}
\left.
\begin{aligned}
X_{\ee}&=X_e^0+\vec U
\end{aligned}
\,\right\rbrace\textrm{ mismatched}.
\end{equation}
where $\vec U$ and $\vec V$ are normally distributed $N_{e}$-dimensional vectors, 
\begin{equation}\label{E:UVDist}
\begin{aligned}
\vec U&\sim \sigma_s N(0,1)\\
\vec V&\sim \sigma_s N(0,1).
\end{aligned}
\end{equation}
Here, $N(0,1)$ is a standard multivariate normal distribution and $\sigma_s=X_e^0/5$ controls the strength of the stimuli. Importantly, this means that each neuron receives a different value of top-down and bottom-up input, in contrast to the previous example (Eq.~\eqref{E:UVhomog} and Figures~\ref{Fig1}--\ref{Fig3}) in which every  neuron in the same excitatory sub-population received the same input. 

Simulating this spiking network model shows that population-averaged firing rates converge to their targets during training on matched stimuli, as expected, but only deviate slightly from their targets in response to a mismatched stimulus (Figure~\ref{Fig4}C). 

We suspected that the deviation of mean excitatory and inhibitory firing rates was small because some neurons increased their firing rates and some neurons decreased their firing rates in response to mismatched stimuli, so the increases and decreases cancelled at the level of population averages. Another way to see this is to note that the expected value of $\vec U$ and $\vec V$ is zero, so the absence of $\vec V$ does not affect the population-averaged value of the inputs and (under a linear approximation) we should not expect a change in mean firing rates by removing $\vec V$. Under this reasoning, the firing rates of individual neurons would still change in response to a mismatched stimulus because individual elements of $\vec V$ are non-zero. This line of reasoning implies that $MSE_{mean}$ should not increase much for a mismatched stimulus, but $MSE_{pop}$ should increase more for a mismatched stimulus. Indeed, this is exactly what we observed in simulations (Figure~\ref{Fig4}D).

In summary, our network model with iSTDP learned to adjust inhibitory weights in such a way to ``match'' or ``cancel'' top-down input with bottom-up input in the sense that the firing rates approach their target rates  in response to matched stimuli after sufficient training. Moreover, the network responded to mismatched stimuli with deviations of the firing rates away from their target values. Note that the deviation of firing rates from their targets is not a consequence of the mismatch alone, but is due to the network being trained on matched stimuli. In this sense, the network is simply detecting deviations of its input patterns from the input patterns on which it was trained. 

\subsection{A lack of detectable prediction errors after training with time-varying stimuli}

While instructive, the examples above were restricted to input patterns that were held fixed during training. In other words, the network only learned to associate {\it one} bottom-up input, $U$, with {\it one} top-down input, $V$ (as schematized in Figure~\ref{Fig3}D). Since animals are exposed to multiple stimuli, a more realistic model would be trained on multiple pairings of top-down and bottom-up inputs. For example, in the visuomotor system, head motion (which we can interpret as top-down input, $V$) is coupled with movement of an animal's visual stimulus (which we can interpret as bottom-up input, $U$). But head motion varies in direction and speed, and the movement of a visual scene covaries with it. Prediction errors arise whenever the learned covariation between head motion and visual stimulus is violated, {\it i.e.}, whenever there is a mismatch between top-down and bottom-up input~\cite{keller2012sensorimotor,attinger2017visuomotor,leinweber2017sensorimotor,jordan2020opposing}. 

 \begin{figure*}
 \centering{
 \includegraphics[width=6in]{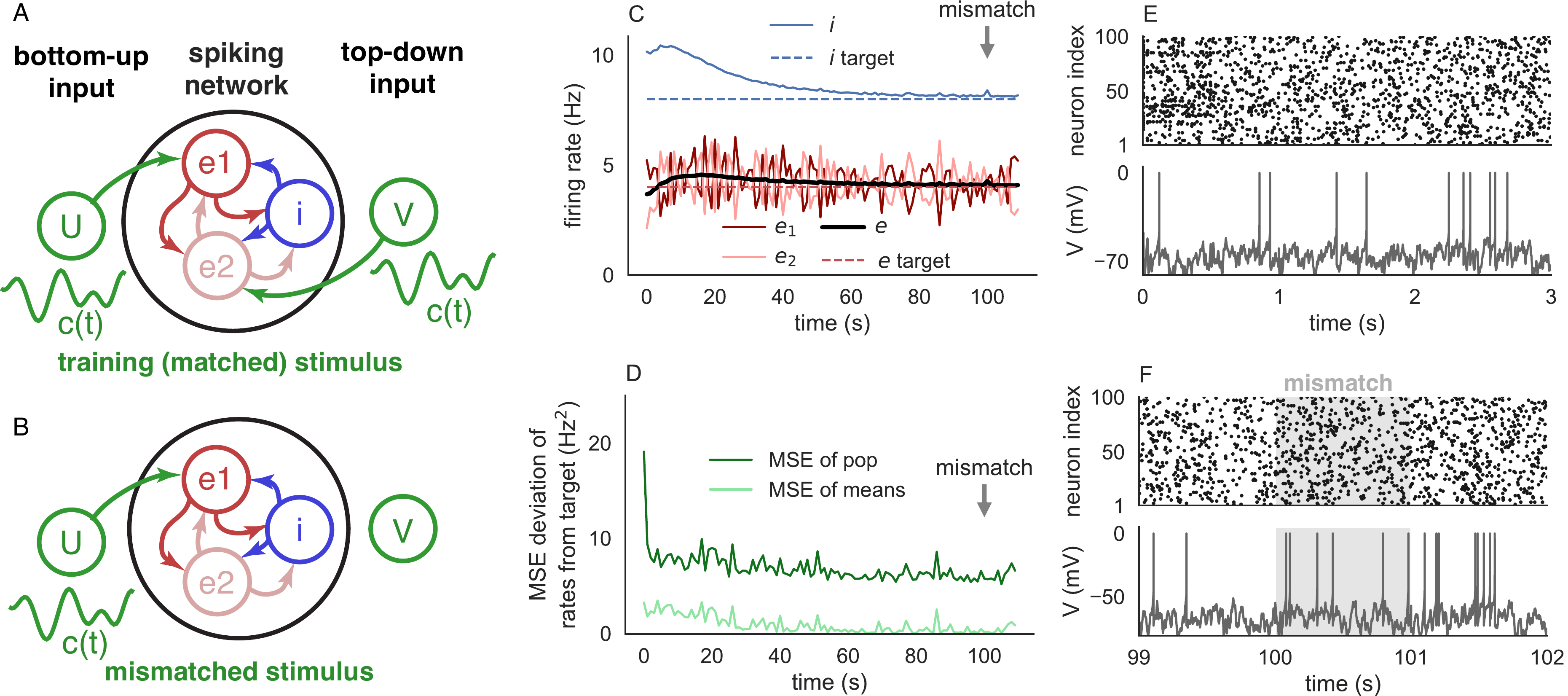}
 }
 \caption{{\bf A lack of detectable prediction errors in a model with time-varying stimuli.} {\bf A,B)} Network schematic. Same as Figure~\ref{Fig1}A except the magnitude of the top-down and bottom-up stimuli were multiplied by the same time-varying signal, $c(t)$.  {\bf C-F)} Same as Figure~\ref{Fig1}C-F except we additionally plotted the mean excitatory firing rates (black curve in C). }
 \label{Fig5}
 \end{figure*}
 
We next considered a simple extension of the first input model from Figures~\ref{Fig1}--\ref{Fig3} to account for top-down and bottom-up inputs with time-varying intensity. Specifically, the excitatory neurons were again broken into two sub-populations, $e_1$ and $e_2$. During training, each neuron in populations $\ee_1$ and $\ee_2$ received external stimuli of the form (Figure~\ref{Fig1}A)
\begin{equation}\label{E:Xmatchedt}
\left.
\begin{aligned}
X_{\ee_1}&=X_e^0+c(t)U\\
X_{\ee_2}&=X_e^0+c(t)V
\end{aligned}
\,\right\rbrace\textrm{ matched}
\end{equation}
where $c(t)$ is a scalar time-series that changes on each trial. Specifically, $c(t)$ is drawn independently from a uniform distribution on $[0,2]$ at the start of each 1s trial.  Hence, the expected value of $c(t)$ is $1$ and therefore, the expected values of $X_{\ee_1}$ and $X_{\ee_2}$ are the same as in the example from Figures~\ref{Fig1}--\ref{Fig3}, but they vary around this expectation across time. 
We used similar top-down and bottom-up, but needed to make the inputs weaker to avoid very large rate deviations, 
\begin{equation}\label{E:UVhomogt}
\begin{aligned}
U&=X_e^0/20\\
V&=-X_e^0/20.
\end{aligned}
\end{equation}
Hence, bottom-up input, $c(t)U$, is matched by top-down input, $c(t)V$, during training. 
After training on matched stimuli, we again modeled mismatched stimuli  by the absence of top-down input
\begin{equation}\label{E:Xmismatchedt}
\left.
\begin{aligned}
X_{\ee_1}&=X_e^0+U\\
X_{\ee_2}&=X_e^0. 
\end{aligned}
\,\right\rbrace\textrm{ mismatched}.
\end{equation}
The input to $e_1$ is not out of the ordinary during a mismatched stimulus (it corresponds to the value when $c(t)=1$ is equal to its expectation) and the input to $e_2$ is not out of the ordinary either (it corresponds to the value when $c(t)=0$), the joint value of the inputs to $e_1$ and $e_2$ together is out of the ordinary because the inputs are not matched (see Figure~\ref{Fig5}A for a schematic).

We reasoned that if our iSTDP rule could learn the relationship between top-down and bottom-up input during training, then it would detect the mismatch between them by evoking a larger deviation of firing rates from their targets. In other words, the network should detect the out-of-distribution input represented by a mismatch. 
However, our spiking network simulations contradicted this prediction. Firing rates deviated from the targets even during matched stimuli and the deviation in response to a mismatched stimulus was similar in magnitude (Figure~\ref{Fig5}B--F). Hence, the the response to a mismatched stimulus was not detectable in the sense that it could not be distinguished from the response to matched stimuli.

\subsection{A mean-field explanation for the absence of mismatch responses after training on time-varying inputs.}

\begin{figure}
 \centering{
 \includegraphics[width=2.25in]{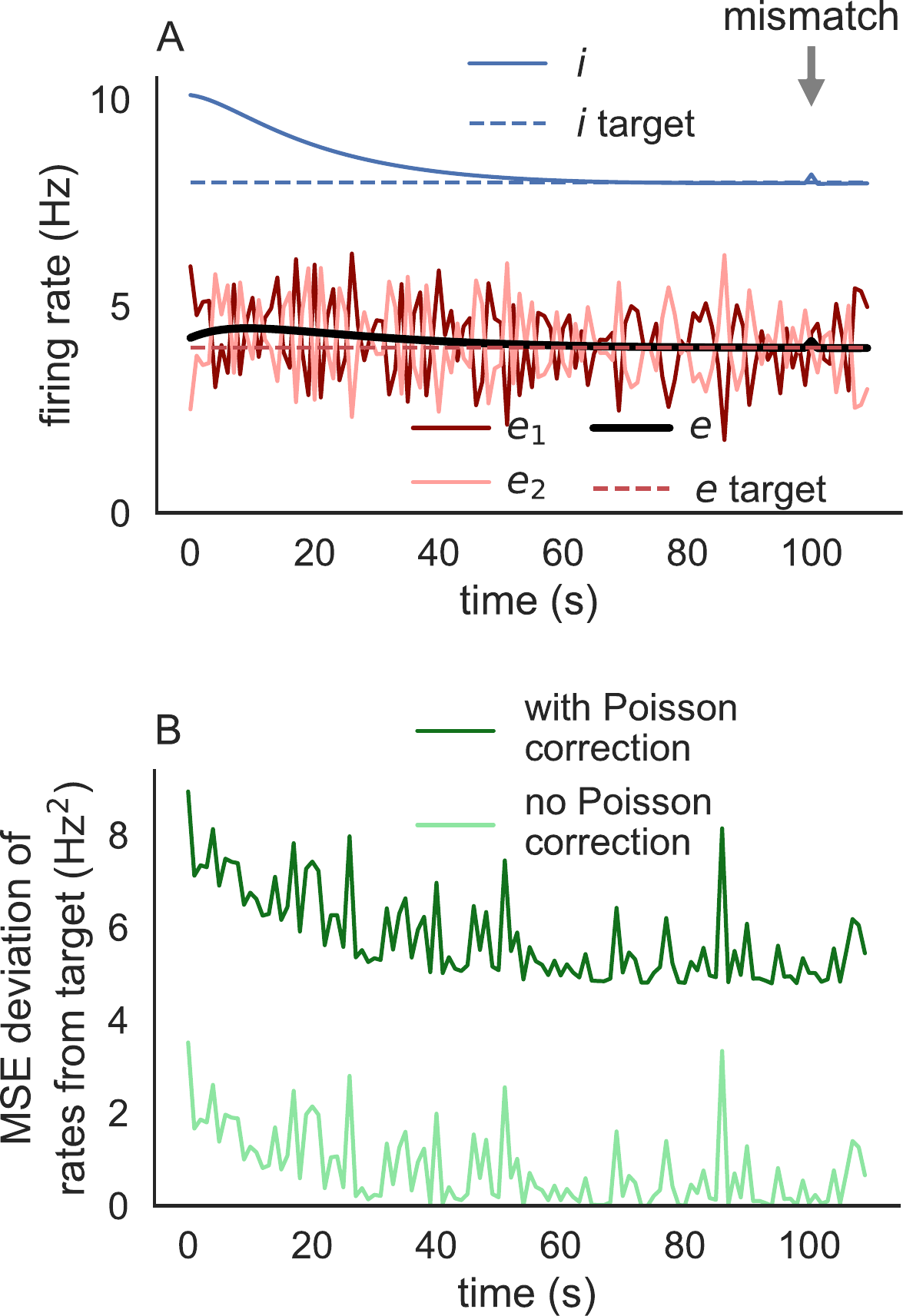}
 }
 \caption{{\bf Mean-field rate model with time-varying stimuli.} {\bf A,B)} Same as Figure~\ref{Fig2} except using the time-varying stimuli from Figure~\ref{Fig5}. }
 \label{Fig6}
 \end{figure}

We now return to our mean-field theory to better understand why we do not see mismatch responses after training on time-varying inputs, but we do see them after training on time-constant inputs. We first simulated dynamical rate model from Eqs.~\eqref{E:drdt}--\eqref{E:fIcurve} with the time-dependent stimuli defined by Eqs.~\eqref{E:Xmatchedt}--\eqref{E:Xmismatchedt}. As above, the dynamical mean-field rate model captured the general trends from the spiking network simulations (compare Figure~\ref{Fig6}A,B to Figure~\ref{Fig5}C,D). 
Eq.~\eqref{E:rqss} for the quasi steady-state firing rates generalizes to 
\begin{equation}\label{E:rqsst}
{\bf r}(W)=\left[D-W\right]^{-1}\vvec X(t)=A\vvec X(t)
\end{equation}
An assumption underlying Eq.~\eqref{E:rqsst} is that $X(t)$ changes more slowly than the timescales ($\tau_a$ for $a=e,i$) at which firing rates evolve. This assumption is valid in our case because $\vvec X(t)$ switches every 1s while $\tau_a\le 6$ms. 

\begin{figure*}
 \centering{
 \includegraphics[width=6in]{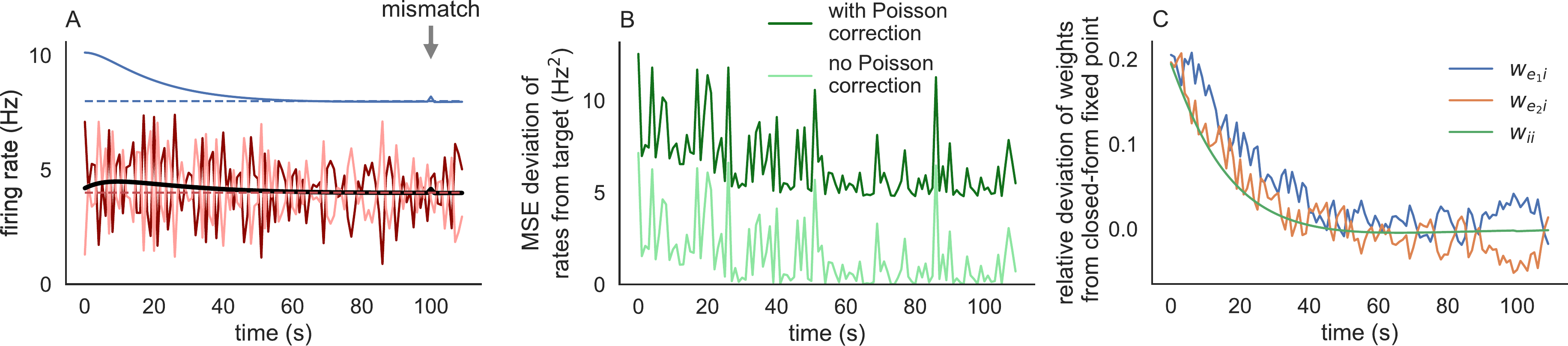}
 }
 \caption{{\bf Slow dynamics captured by a separation of timescales in a model with time-dependent stimuli.}  {\bf B-C)} Same as Figure~\ref{Fig3} except using the time-varying stimuli from Figure~\ref{Fig5}B-C. {\bf D)} Same as Figure~\ref{Fig5}D except time-dependent stimuli during training are represented by multiple dots (each one representing the inputs on one trial) and the mean is represented by a purple x.}
 \label{Fig7}
 \end{figure*}

Now we can transition to the slower timescale dynamics of $W$ by re-writing Eqs.~\eqref{E:dwrs} as
\begin{equation}\label{E:dwrst}
\begin{aligned}
\frac{dW}{dt}&=\left[\begin{array}{ccc} 0 &  0 & -\eta_e(r_{e_1}(t)-r^e_0)r_i(t)\\ 0 &  0 & -\eta_e(r_{e_2}(t)-r^e_0)r_i(t)\\ 0 & 0 & -\eta_i(r_i(t)-r^i_0)r_i(t)\end{array}\right]\\
\vvec r(t)&=\left[\begin{array}{c}r_{e_1}(t)\\ r_{e_2}(t)\\ r_i(t)\end{array}\right]=\left[D-W\right]^{-1}\vvec X(t)=A\vvec X(t)
\end{aligned}
\end{equation}
where we have only added the explicit time-dependence. Simulating this system shows general agreement with the trends from the spiking networks simulations and the dynamical mean-field model (Figure~\ref{Fig7}A,B, compare to Figure~\ref{Fig5}C,D and Figure~\ref{Fig6}A,B). 

Due to the time-dependence of $\vvec X(t)$ in the current example, Eqs.~\eqref{E:dwrst} do not have a fixed point, so we cannot proceed directly with the fixed point analysis from above. 
To perform a fixed point analysis on $W$,  we must assume that plasticity is slower than the stimulus, {\it i.e.}, that $W(t)$ changes much more slowly than $\vvec X(t)$. This assumption is valid for our simulations and even more so for biological neural circuits. Under this assumption, the slow timescale dynamics of $W$ evolve based on the mean value of $\vvec X(t)$. Specifically, we can use the approximation
\begin{equation}\label{E:dwrstm}
\begin{aligned}
\frac{dW}{dt}&=\left[\begin{array}{ccc} 0 &  0 & -\eta_e(\overline r_{e_1}-r^e_0)\overline{r_i}\\ 0 &  0 & -\eta_e(\overline r_{e_2}-r^e_0)\overline{r_i}\\ 0 & 0 & -\eta_i(\overline r_i-r^i_0)\overline{r_i}\end{array}\right]\\
\overline{\vvec r}&=\left[\begin{array}{c}\overline r_{e_1}\\ \overline r_{e_2}\\ \overline r_i\end{array}\right]=\left[D-W\right]^{-1}\overline{\vvec X}=A\overline{\vvec X}
\end{aligned}
\end{equation}
where
\[
\overline{\vvec X}=E_t[\vec X(t)]
\]
and $E_t$ denotes the expectation over time during training, {\it i.e.}, during matched stimuli. 

During training (for matched stimuli), we have from Eq.~\eqref{E:Xmatchedt} that 
\begin{equation}\label{E:Xmatched}
\vvec X^m(t)=\left[\begin{array}{c}X_e^0 +c(t)U\\ X_e^0 +c(t)V\\ X_i^0\end{array}\right]
\end{equation}
Since $E_t[c(t)]=1$, we have that 
\begin{equation}\label{E:Xbar}
\overline{\vvec X}=\left[\begin{array}{c}X_e^0+U\\X_e^0+V\\ X_i^0\end{array}\right]
\end{equation}
which is the same as the model from Figures~\ref{Fig1}--\ref{Fig3}. Hence, under this approximation, $W$ should converge to the same fixed point in Eq.~\eqref{E:wFP}. Notably, this implies that the time-averaged rates should be equal to the target rates, $\overline{\vvec r}=\vvec r^0$. 
As predicted, simulations show that average firing rates are close to their targets (Figure~\ref{Fig7}A) and the weights do converge to the given fixed point with the addition of some noise (Figure~\ref{Fig7}C) coming from the noisy time-dependence of $\vvec X(t)$ and $\vvec r(t)$. 

Therefore, the state of the network (as represented by $W$) after training is similar for the networks with time-constant and time-dependent stimuli. As a result, the deviation, $d\vvec r(t)$, of the firing rates from their targets on any given trial takes the same form derived in Eq.~\eqref{E:dr},
\begin{equation}\label{E:drt}
\begin{aligned}
d\vvec r(t)&:=\vvec r(t)-\vvec r^{0}\\
&=A^0\vvec X(t)-\vvec r^{0}\\
&=A^0(\vvec X(t)-\overline{\vvec X})\\
&=A^0d\vvec X(t)
\end{aligned}
\end{equation}
where $d\vvec X(t)=\vvec X(t)-\overline{\vvec X}$ is the deviation of the stimulus from the mean value it takes during training and $A^0$ is the fixed point of $A=[D-W]^{-1}$ after training (see Eq.~\eqref{E:Wr0} and surrounding discussion). This conclusion assumes that the  mean-field approximation in Eq.~\eqref{E:rqsst} is approximately accurate or, more specifically, that the firing rate response to a perturbation is approximately a linear function of the input perturbation. This, in turn, requires that the input perturbation is not too strong.

 \begin{figure}[h]
 \centering{
 \includegraphics[width=3.25in]{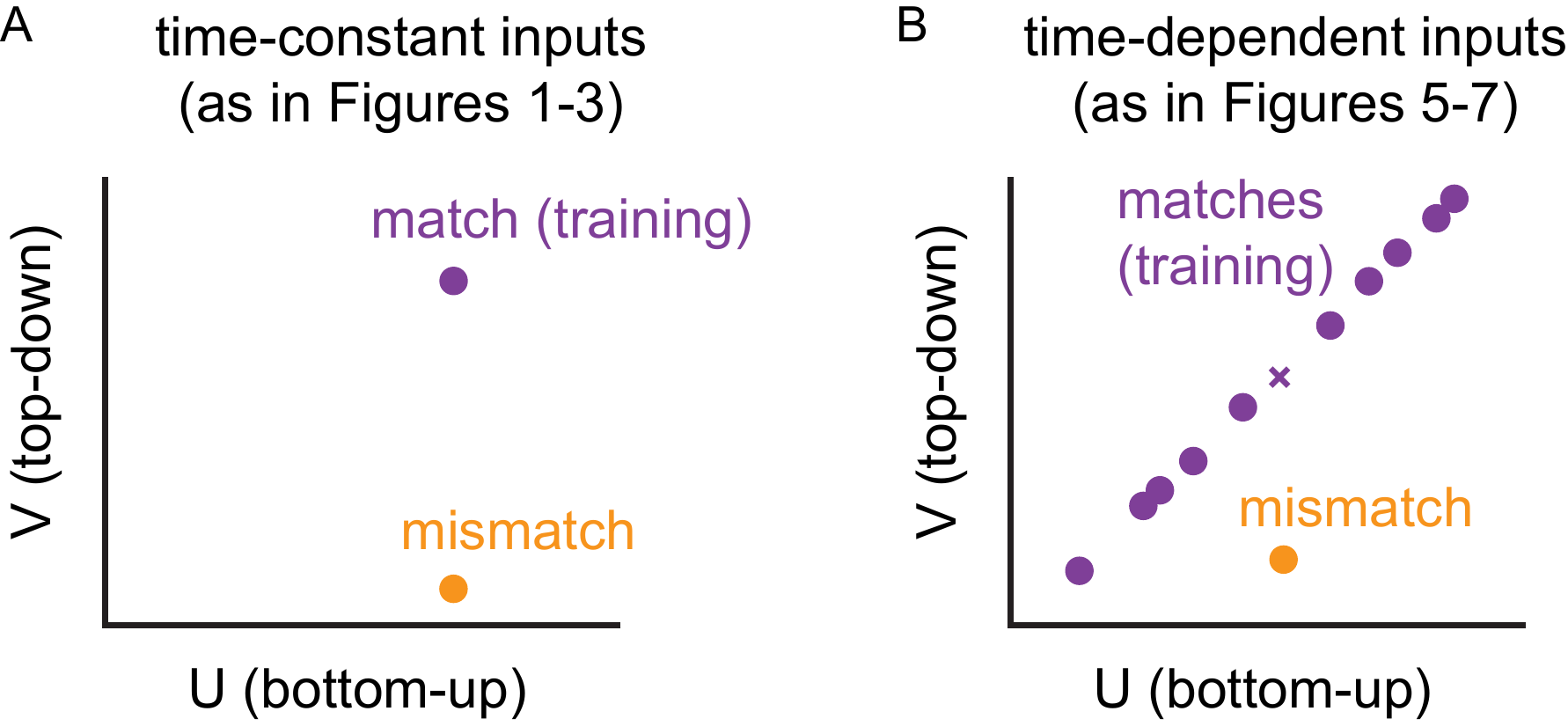}
 }
 \caption{{\bf Schematic illustrating why mismatch responses are detectable after training on time-constant, but not time-dependent stimuli.} {\bf A)} Schematic representing inputs to the network in a model with time-constant stimuli. Training stimuli  occupy a single point in $(U,V)$ space (purple dot). The deviation of firing rates from their targets on any particular trial is approximately proportional to the distance of the input from its value during training (Eq.~\eqref{E:dr}). Since the mismatch stimulus (orange dot) is far from the matched, training stimulus, firing rates deviate from their target in response to the mismatched stimulus (as seen in Figures~\ref{Fig1}--\ref{Fig3}). {\bf B)} Schematic representing inputs to the network in a model with time-varying stimuli. Training stimuli (purple dots) vary in $(U,V)$ space along a predictable line. The mismatched stimulus lies far from this line. However,  the deviation of firing rates from their targets on any particular trial is approximately proportional to the distance of the input from its {\it mean} value during training (Eq.~\eqref{E:dr}). Since the distance between the mismatch input (orange dot) and the mean training stimulus (purple x) is similar to the typical distance between the individual training stimuli (purple dots) and the mean training stimulus (purple x), the deviation of the firing rates from their targets is similar for matched and mismatched stimuli.   }
 \label{Fig8}
 \end{figure}

As a heuristic, we can ignore the effect of $A^0$ in Eq.~\eqref{E:drt} and make the approximation that $d\vvec r(t)$ is larger whenever $d\vvec X(t)$ is larger. In other words,
\begin{equation}\label{E:AX0approx}
\begin{aligned}
\|d\vvec r(t)\|&=\|A^0 d\vvec X(t)\|\\
&\approx \|A^0\|\|d\vvec X(t)\|\\
&\propto \|d\vvec X(t)\|.
\end{aligned}
\end{equation}
where $\|A^0\|$ denotes the induced Euclidean norm on $A^0$. 
In other words, stimuli that are further from the mean training stimuli evoke larger firing rates. Note that we necessarily have $\|A^0 d\vvec X(t)\|\le  \|A^0\|\|d\vvec X(t)\|$, so this assumption is saying that $\|A^0 d\vvec X(t)\|$ is not much smaller than $\|A^0\|\|d\vvec X(t)\|$. 
This approximation assumes that $d\vvec X(t)$ is not close to being orthogonal to the rows of $A^0$. 

During matched stimuli, combining Eqs.~\eqref{E:Xmatched} and \eqref{E:Xbar} gives the perturbation for training stimuli 
\[
d\vvec X^m(t)=\left[\begin{array}{c}(1-c(t))U\\ (1-c(t))V\\ 0 \end{array}\right].
\]
Since $|U|=|V|$, we have 
\[
\begin{aligned}
\left\|d\vvec X^m(t)\right\|^2&=2(1-c(t))^2 |V|\\
&=2u^2(t)|V|
\end{aligned}
\]
where $u(t)=1-c(t)$ is uniformly distributed on $[-1,1]$. Hence, the squared distance of $\vvec X(t)$ from its mean varies between $0$ and $2|V|$. During the mismatched stimulus, we have from Eq.~\eqref{E:Xmismatchedt}, that 
\[
\vvec X^{mm}=\left[\begin{array}{c}X_e^0 +U\\ X_e^0\\ X_i^0\end{array}\right]
\]
Combining this with Eq.~\eqref{E:Xbar} shows that, during a mismatched stimulus, the input perturbation is 
\[
d\vvec X^{mm}=\left[\begin{array}{c}0\\ -V\\ 0\end{array}\right]
\]
and therefore 
\[
\left\|d\vvec X^{mm}\right\|^2=|V|.
\]
Hence, the deviation of the external input, $\vvec X(t)$, from its mean value during training is similar in magnitude during matched and mismatched stimuli. As a result, the deviation of the firing rates from their targets is also similar during matched and mismatched stimuli, so the mismatch is not detectable based on the deviation of firing rates from their targets alone.

This intuition, and how it differs from the time-constant model of Figures~\ref{Fig1}--\ref{Fig3}, is illustrated in Figure~\ref{Fig8}.  For the model with  time-constant inputs, there is only one stimulus during matched, training trials (Figure~\ref{Fig8}A, purple dot). Since the mismatch stimulus is far from this matched stimulus, the firing rate deviates from its target in response to the mismatched stimulus (as demonstrated in Figures~\ref{Fig1}--\ref{Fig3}). For the model with time-varying stimuli,  there are multiple training stimuli that lie along a line (Figure~\ref{Fig8}B, purple dots). While the mismatch stimulus is clearly away from this line (Figure~\ref{Fig8}B, orange dot), the deviation of the firing rates from their targets is approximately proportional to how far an input is from the {\it mean} training stimulus (Figure~\ref{Fig8}B, purple x). Since this distance is similar for the mismatch stimulus and a typical training stimulus, the deviation of the firing rates from their targets is also similar during matched and mismatched stimuli (as demonstrated in Figures~\ref{Fig5}--\ref{Fig7}). While this intuition might seem obvious in hindsight, the complexity of dynamics in recurrent spiking neural network models can make this conclusion difficult to foresee without the benefit of the mean-field analysis provided here.

\begin{figure*}
 \centering{
 \includegraphics[width=6in]{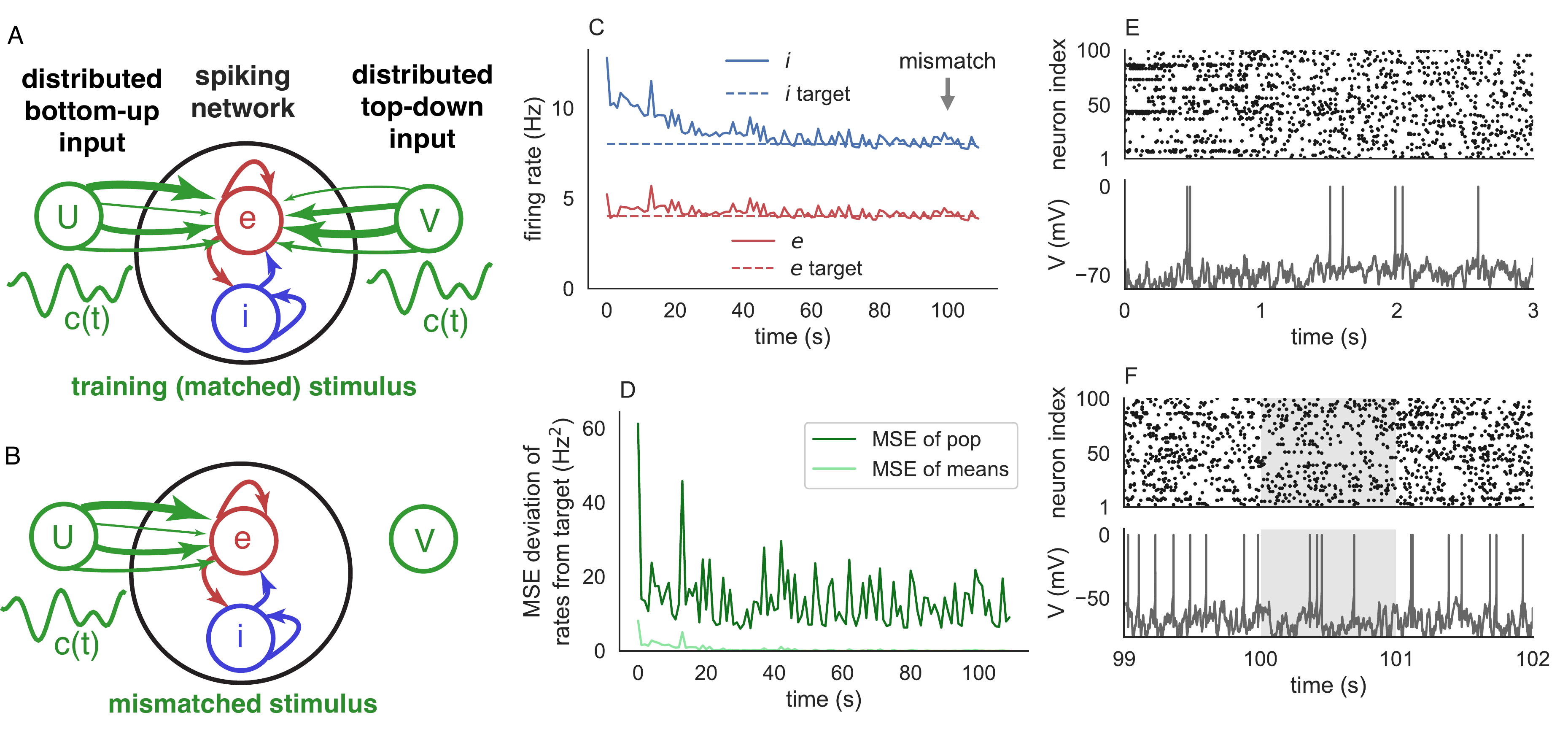}
 }
 \caption{{\bf Prediction errors after training on distributed time-dependent inputs.} Same as Figure~\ref{Fig4} except bottom-up and top-down inputs were time-dependent, as described by Eqs.~\eqref{E:XDistmatchedt}--\eqref{E:XDistmismatchedt}.}
 \label{Fig9}
 \end{figure*}

For the sake of completeness, we also considered a model with distributed, time-varying stimuli. Specifically, we combined the time-varying stimuli from the example in Figure~\ref{Fig7} with the distributed stimuli from the example in Figure~\ref{Fig4} to get inputs of the form (Figure~\ref{Fig9}A,B)
\begin{equation}\label{E:XDistmatchedt}
\left.
\begin{aligned}
X_{\ee}&=X_e^0+c(t)\vec U+c(t)\vec V\\
\end{aligned}
\,\right\rbrace\textrm{ matched}
\end{equation}
and
\begin{equation}\label{E:XDistmismatchedt}
\left.
\begin{aligned}
X_{\ee}&=X_e^0+\vec U
\end{aligned}
\,\right\rbrace\textrm{ mismatched}.
\end{equation}
where $c(t)$ is a scalar drawn from a uniform distribution on $[0,2]$ on each trial, and $\vec U$ and $\vec V$ are normally distributed $N_{e}$-dimensional vectors as in Eq.~\eqref{E:UVDist}. Unsurprisingly, given the failure on the simpler example discussed above, the spiking network model did not produce an easily detectable response to mismatched stimuli (Figure~\ref{Fig9}C-F). Specifically, the deviation of the firing rates away from their targets was similar in matched and mismatched trials (Figure~\ref{Fig9}C,D). 


\subsection{How do our conclusions generalize to other network models?}

The mean-field analysis above relied on several assumptions that were used to derive approximations. This raises the question of how general our conclusions are. Specifically, for which network models does the argument above imply an absence of noticeable mismatch responses? To answer this question, we can distill the argument above into three fundamental assumptions: 
\begin{enumerate}
\item The linear approximation in Eq.~\eqref{E:drt} should be approximately accurate,
 \[
 d\vvec r(t)\approx A^0 d\vvec X(t)
 \]
 While this assumption is strong, it should be satisfied when $d\vvec X(t)$ is sufficiently small. In addition, balanced excitation and inhibition linearize the firing rate responses of networks to external input~\cite{vanVreeswijk:1998uz,Rosenbaum2014,lim:2014,landau2016impact,ebsch2018imbalanced,ahmadian2021dynamical}, so this assumption should hold in networks with balanced excitation and inhibition, which is encouraged by inhibitory synaptic plasticity~\cite{Vogels2011,Hennequin2017,baker2020nonlinear,akil2021balanced}.
\item The approximation in Eq.~\eqref{E:AX0approx} should be accurate, specifically
\[
\|A^0 d\vvec X(t)\|\approx \|A^0\|\|d\vvec X(t)\|
\]
which requires that $d\vvec X(t)$ not be close to orthogonal to the rows of $A^0$. 
\item The magnitude of the input perturbations for a mismatched stimulus should be similar to a typical value during matched stimuli,
\[
\|d\vvec X^{mm}\|\approx \|d\vvec X^{m}(t)\| 
\]
\end{enumerate}
In general, if a model satisfies these three assumptions then $d\vvec r(t)$ is similar in magnitude during matched and mismatched stimuli. 
Note that these assumptions are sufficient, but not necessary for a lack of mismatch responses. For example, if assumption 1 is violated because the rate perturbations are nonlinear, then the nonlinear model might still not compute mismatch responses.

Strictly speaking, assumption 2 is stronger than needed. Instead, we only need that the relationship between $A^0$ and $d\vvec X$ is similar for matched and mismatched stimuli, {\it i.e.}, that
\[
\frac{\|A^0 d\vvec X^m(t)\| }{\|A^0\|\|d\vvec X^m(t)\|}\approx \frac{\|A^0 d\vvec X^{mm}\| }{\|A^0\|\|d\vvec X^{mm}\|}
\]
which is a weaker assumption because it allows for $d\vvec X$ to be aligned with the rows of $A^0$ so long as the alignment is similar for matched and unmatched stimuli. 

For our examples in which the network is trained on time-constant input (Figures~\ref{Fig1}--\ref{Fig4}), we have that $\vvec X^m(t)=\overline{\vvec X}$, so $d\vvec X^m(t)=0$ whereas  $d\vvec X^{mm}\ne 0$, so assumption 3 above is not met. This explains why our examples trained on time-constant were able to produce robust mismatch responses.

In previous work~\cite{hertag2020learning}, a network with homeostatic plasticity successfully computed prediction errors after training on time-varying stimuli. In that work, the weights of the connectivity matrix were carefully chosen so that $A$ was singular and the directions of the input perturbation during matched stimuli (the ``feedback'' stimulus condition) was in the nullspace of $A^0$. See equation 28 in their appendix and note that $A^0$ was called $W$ in their analysis. As a result, the model studied there does not satisfy assumption 2 above. This explains how \cite{hertag2020learning} were able to compute prediction errors with time-varying inputs.

In all of the examples we have considered so far, external input was provided to excitatory neurons only. However, our analysis implies that our overall results should still hold if input is provided to inhibitory neurons as well. Specifically, in Eqs.~\eqref{E:drt} and the surrounding equations and analysis, there is nothing preventing $d\vvec X(t)$ from having a non-zero component for the inhibitory population(s). To verify this prediction, we repeated all of the spiking network simulations (those in Figures~\ref{Fig1}, \ref{Fig4}, \ref{Fig5}, and \ref{Fig9}) in models in which external input was also added to the inhibitory population. Our results show the same overall conclusions for all figures (see Supplementary Materials Section~1 and Supplementary Figures~1--2). Specifically, in all examples, a noticeable mismatch response was observed after training on time-constant inputs, but not after training on time-varying inputs. 

Assumption 3 above implies that mismatch responses could be possible after training on time-varying stimuli if the mismatch stimulus is larger in magnitude than the matched stimuli used during training. While this is not necessarily a surprising finding (a larger stimulus should evoke a larger response), we decided to test it in a simulation. Specifically, we repeated the simulation from Figure~\ref{Fig5}, but we scaled the magnitude of the mismatched input by a factor of six. These simulations confirm that a mismatch response was produced in this case (Supplementary Figure 3). 

In all of the examples above, we considered only a single inhibitory population and at most two excitatory populations. In reality, there are multiple inhibitory neuron subtypes in the cortex and previous work on mismatch responses with inhibitory plasticity accounts for this~\cite{hertag2020learning,hertag2021prediction}. Our analysis above implies that increasing the number of neuron populations alone should not affect our overall conclusions. To test our findings empirically on a model with several neural populations, we performed a simulation that was identical to the simulation in Figure~\ref{Fig5} except we used three inhibitory and three excitatory populations. Consistent with our theoretical predictions, the results were qualitatively similar to those in Figure~\ref{Fig5}: After training on time-dependent stimuli, there was no noticeable deviation of firing rates in response to a mismatched stimulus (see Supplementary Figure 4). 


%

\section{Discussion}

We combined numerical simulations of spiking networks and mean-field rate models with mathematical analysis to evaluate the extent to which homeostatic inhibitory synaptic plasticity can train an unstructured network to compute prediction errors. We found that the networks successfully learn to compute prediction errors when training stimuli are static. Specifically, if top-down and bottom-up inputs are fixed in time during training, then firing rates in the trained network will maintain a baseline firing rates in response to stimuli that match the training stimuli, but firing rates will deviate from their baseline levels in response to mismatched stimuli. This result holds when stimuli are uniform (with each of a few sub-populations receiving homogeneous external input) or when stimuli are distributed (with each neuron receiving distinct, but time-constant levels of external input during training). 

To our surprise, simulations showed that even under a simple model of time-varying stimuli, in which bottom-up and top-down inputs are modulated by the same time-varying factor, the same networks fail to produce reliable mismatch responses after training. Specifically, firing rates deviate from their baseline levels by a similar amount in response to stimuli that are matched (a shared modulation, as in training) or mismatched (one input is modulated differently than the other). We used a mean-field approximation to explain these empirical findings and elucidate a set of conditions under which robust mismatch responses do not occur. Our results therefore help to clarify the extent to which homeostatic inhibitory synaptic plasticity is sufficient to train a network to compute mismatch responses. 

For networks trained on time-varying inputs, our results show a lack of mismatch responses in the sense that firing rates do not deviate from their baseline (when deviation is measured by mean-squared error) more during mismatched inputs than they do for matched stimuli. However, mismatch responses could potentially be detected by some linear projection of the firing rates and this linear projection could be fed as input to a readout neuron that would be able to detect mismatch responses. However, our main goal was to understand the situations under which a natural homeostatic plasticity rule would spontaneously produce elevated responses to mismatched stimuli. Training a separate linear projection is outside the scope of this goal.

Inhibitory homeostatic synaptic plasticity is only one of many homeostatic mechanisms in the brain~\cite{turrigiano2011too}. While homeostatic plasticity is one candidate mechanism for predictive coding, other homeostatic mechanisms could play a role as well. Future work should consider the potential role of other homeostatic mechanisms in predictive coding and mismatch detection. 

Previous work~\cite{hertag2020learning,hertag2021prediction} found that networks with homeostatic plasticity {\it can} learn to compute mismatch responses in models with time-varying stimuli that are similar to the time-varying stimuli that we used (in the cases where our networks failed). They used a more biologically detailed network model with multiple inhibitory subtypes and multi-compartment excitatory neurons. Importantly, connectivity in their model was constrained so that matched stimuli were in the nullspace of the effective connectivity matrix ($A$ in our work, $W$ in theirs). Our theoretical analysis agrees with their analysis showing that this assumption is necessary for their overall results. We additionally provided a set of conditions under which more general classes of models will not produce robust mismatch responses, which generalizes some of the theoretical results in \cite{hertag2020learning} to more general classes of networks. The requirement that matched stimuli are in the nullspace of the effective connectivity matrix is a strong assumption because it implies that the connectivity matrices must be precisely tuned. Moreover, the dimension of the nullspace of the connectivity matrix must match the dimensionality of the training stimuli, which could make it difficult to train a network to maintain baseline firing rates on a higher dimensional space of training stimuli. 


Our study and the previous work described above~\cite{hertag2020learning,hertag2021prediction} incorporates homeostatic synaptic plasticity, but does not account for any other of the wide variety of synaptic plasticity rules observed in neural recordings. Other work has shown that predictive coding can be learned in carefully constructed networks using  learning rules that are not exclusively homeostatic~\cite{bogacz2017tutorial}. Indeed, our approach of learning prediction errors in unstructured, randomly connected networks could potentially be made successful if the target rates, $r_0^a$, were effectively modulated by the top-down or bottom-up input. Future work should consider the possibility of learning prediction errors in unstructured, random networks by combining these approaches.


\section{Materials and Methods}

All simulations were performed by numerically solving the corresponding differential equations using  the forward Euler method in custom written Python code. Code to produce all figures can be found at \texttt{https://github.com/RobertRosenbaum/PCISP}.

For spiking network simulations (Eqs.~\eqref{E:dV}--\eqref{E:dx}; Figures~\ref{Fig1}, \ref{Fig4}, and \ref{Fig5}) and mean-field rate network simulations (Eqs.~\eqref{E:drdt}--\eqref{E:dwaidt}; Figures~\ref{Fig2} and \ref{Fig6}) we used a time step size of $dt=0.1$ms. For the slow-timescale model (Eqs.~\eqref{E:dwrstm}; Figures~\ref{Fig3} and \ref{Fig7}) we used a time step size of $dt=1$s. 

For all spiking network simulations (Eqs.~\eqref{E:dV}--\eqref{E:dx}; Figures~\ref{Fig1}, \ref{Fig4}, and \ref{Fig5}), we used $N_e=4000$ and $N_i=1000$ excitatory and inhibitory neurons. All neurons were connected with probability $p_{ee}=p_{ei}=p_{ie}=p_{ii}=0.1$. Connected neurons had initial synaptic weights $j_{ee}=7.07$mV/ms, $j_{ei}=-49.5$mV/ms, $j_{ie}=31.8$mV/ms, and $j_{ii}=-70.7$mV/ms. EIF neuron parameters were $\tau_m=15$ms, $E_L=-72$mV, $V_{re}=-73$mV, $D_T=2$mV, $V_T=-55$mV, $V_{th}=0$mV, and a reflecting lower boundary on the membrane potential was placed at $V_{lb}=-80$mV to approximate an inhibitory reversal potential. Synaptic timescales were $\tau_e=6$ms and $\tau_i=4$ms. Baseline external input to excitatory and inhibitory neurons was $X_e^0=42.4$mV and $X_i^0=28.3$mV. Parameters for the inhibitory plasticity rule were $\eta_e=56.6$mV, $\eta_i=28.3$mV, and $\tau_{STDP}=200$ms with target rates at $r_0^e=4$Hz and $r_0^i=8$Hz. 
For mean-field rate network simulations (Eqs.~\eqref{E:drdt}--\eqref{E:dwaidt}; Figures~\ref{Fig2} and \ref{Fig6}), we used a gain of $g=0.001$ms/mV, which was derived by simulating the spiking network model without plasticity and then fitting the f-I curve $ r=f( I)=g I H( I)$ (where $H$ is the Heaviside step function) to the time-averaged firing rates and input currents of all neurons in the simulation. 
Learning rates for rate network simulations were $\eta_e=8944$mV and $\eta_i=4472$mV. All other parameters were the same as those used in spiking network simulations or their derivations are given in Results. Python code to simulate the networks reproduce the figures can be found on the last  author's academic webpage.

\subsection{Derivation of Eq.~\eqref{E:MSEPoisson} for $MSE_{Poisson}$.}

Here, we derive Eq.~\eqref{E:MSEPoisson} for $MSE_{Poisson}$. Consider a population of $N$ neurons divided into $M$ sub-populations  where sub-population $a$ contains $N_a$ neurons for $a=1,\ldots, M$ ($M=3$ and $a=e_1,e_2,i$ for the models considered in this paper). Assume that each neuron in population $a$ spikes like a Poisson process with a rate of $r_a$. Let $n^a_j$ be the number of spikes emitted by neuron $j=1,\ldots, N_a$ in population $a=1,\ldots, M$ during a time interval of duration $T$ and let 
\[
r^a_j=\frac{n^a_j}{T}
\] 
be the sample firing rate of neuron $j$. Then each $n^a_j$ has expectation and variance 
\[
E[n^a_j]=\var(n^a_j)=r_a T
\] 
so each sample rate has  expectation  
\[
\begin{aligned}
E[r^a_j]&=E\left[\frac{n^a_j}{T}\right]
=r_a
\end{aligned}
\] 
and variance 
\[
\begin{aligned}
\var(r^a_j)=\var\left(\frac{n^a_j}{T}\right)
=\frac{r_a}{T}.
\end{aligned}
\]
Now suppose we have a target rates of $r_a^0$ for each neuron in population $a$ and we would like to compute the population-wide MSE deviation of the sample rates from their targets. This can be written as 
\[
\begin{aligned}
MSE_{pop}&=\frac{1}{N}\sum_{j=1}^N (r_j-r^0_j)^2\\
&=\frac{1}{N}\sum_{a=1}^M\sum_{j=1}^{N_a} (r^a_j-r_a^0)^2\\
&=\sum_{a=1}^Mq_a\frac{1}{N_a}\sum_{j=1}^{N_a} (r^a_j-r_a^0)^2
\end{aligned}
\]
where $q_a=N_a/N$ is the proportion of neurons in population $a$,  $r_j$ is the sample rate, and $r^0_j$ is the rate parameter for neuron $j=1,\ldots, N$. The inner sum can be written as
\[
\begin{aligned}
\frac{1}{N_a}\sum_{j=1}^{N_a} &(r^a_j-r_a^0)^2=(r_a^0-r_a)^2\\
&+\frac{1}{N_a}\sum_{j=1}^{N_a}(r^a_j-r_a)^2-2(r_a^0-r_a)(r^a_j-r_a).
\end{aligned}
\]
The first term in the sum is  the sample variance of $r^a_j$, so 
\[
\frac{1}{N_a}\sum_{j=1}^{N_a}(r^a_j-r_a)^2\approx \var(r^a_j)=\frac{r_a}{T}.
\]
when $N_a$ is large. The last term in the sum can be ignored when $N_a$ is large because 
\[
\begin{aligned}
\frac{1}{N_a}\sum_{j=1}^{N_a}(r_a^0-r_a)(r^a_j-r_a)&=(r_a-r_a^0)\left(r_a-\frac{1}{N_a}\sum_{j=1}^{N_a} r^a_j\right)\\
&\approx 0
\end{aligned}
\]
since $r_a$ is the expected value of $r^a_j$. Putting this altogether gives 
\[
\begin{aligned}
MSE_{pop}&\approx \sum_{a=1}^M q_a \left[(r_a-r_a^0)^2+\frac{r_a}{T}\right]\\
&=MSE_{mf}+\frac{1}{T}\sum_{a=1}^M q_a r_a
\end{aligned}
\]
where
\[
MSE_{mf}=\sum_{a=1}^M q_a (r_a-r^0_a)^2
\]
is the mean-field MSE defined in Eq.~\eqref{E:MSEmf}. This calculation motivates the definition of the Poisson-corrected MSE,
\[
MSE_{Poisson}=MSE_{mf}+\sum_{a=1}^M \frac{q_a r_a}{T}
\]
as defined in Eq.~\eqref{E:MSEPoisson}. Specifically, our calculations above show that $MSE_{Poisson}$ approximates the population-level MSE ({\it i.e.}, $MSE_{pop}$) that would be produced if all of the spike trains in each sub-populations were Poisson processes. The approximation becomes exact as $N_a\to\infty$.

\section*{Declarations}

\subsection*{Funding and/or conflicts of interest.} 

 This work was supported by US National Foundation of Science grants NSF-DMS-1654268 and NSF NeuroNex DBI-1707400, and the Air Force Office of Scientific Research (AFOSR) under award number FA9550-21-1-0223. The  authors  have  no  conflicts  of  interest  to  disclose.


\includepdf[pages=-,pagecommand={},width=\textwidth]{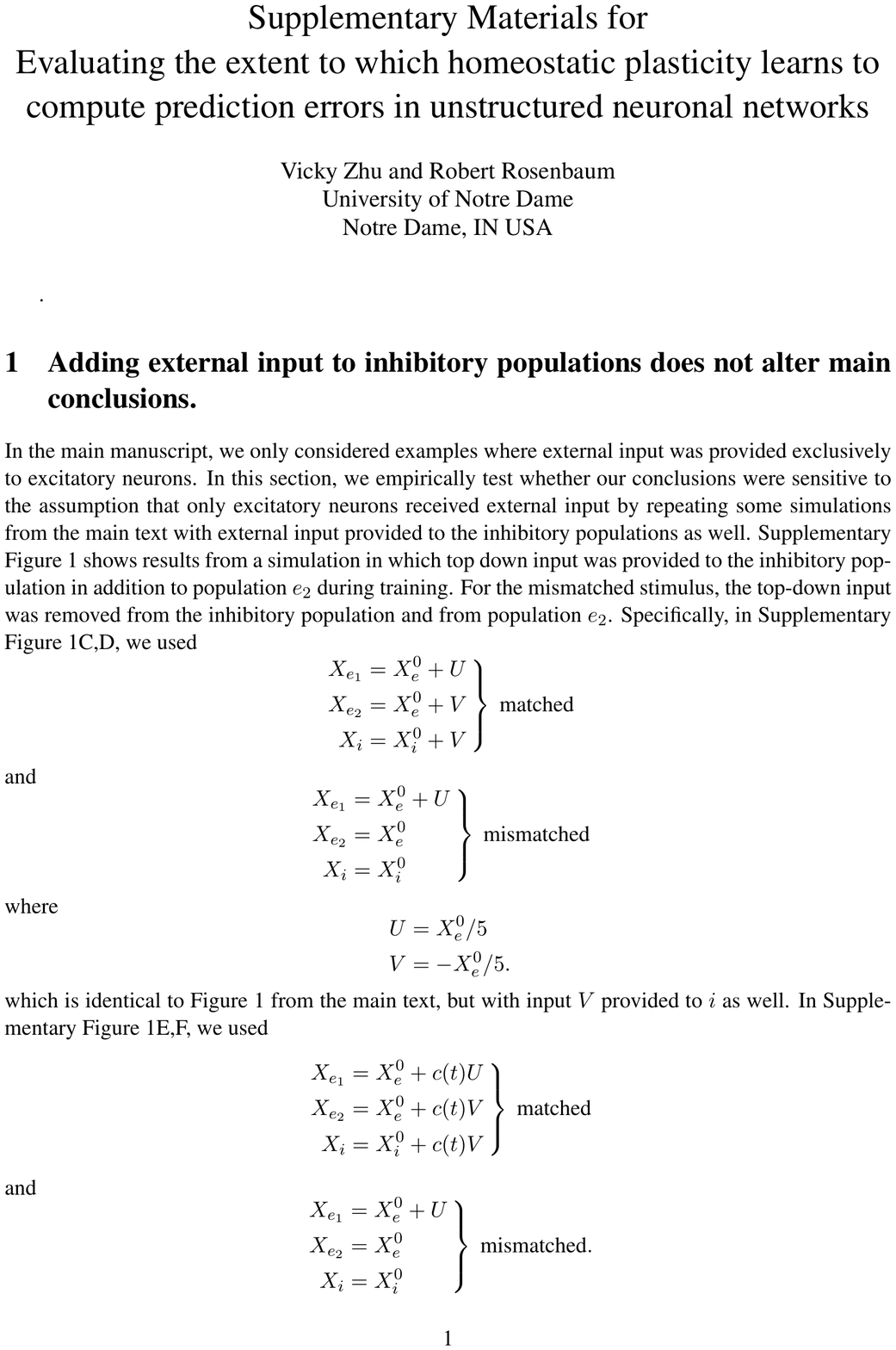}


\begin{thebibliography}{10}

\bibitem{keller2012sensorimotor}
G.~B. Keller, T.~Bonhoeffer, and M.~H{\"u}bener.
\newblock Sensorimotor mismatch signals in primary visual cortex of the
  behaving mouse.
\newblock {\em Neuron}, 74(5):809--815, 2012.

\bibitem{leinweber2017sensorimotor}
M.~Leinweber, D.~R. Ward, J.~M. Sobczak, A.~Attinger, and G.~B. Keller.
\newblock A sensorimotor circuit in mouse cortex for visual flow predictions.
\newblock {\em Neuron}, 95(6):1420--1432, 2017.

\bibitem{attinger2017visuomotor}
A.~Attinger, B.~Wang, and G.~B. Keller.
\newblock Visuomotor coupling shapes the functional development of mouse visual
  cortex.
\newblock {\em Cell}, 169(7):1291--1302, 2017.

\bibitem{von1867handbuch}
H.~Von~Helmholtz.
\newblock {\em Handbuch der physiologischen Optik: mit 213 in den Text
  eingedruckten Holzschnitten und 11 Tafeln}, volume~9.
\newblock Voss, 1867.

\bibitem{keller2018predictive}
G.~B. Keller and T.~D. Mrsic-Flogel.
\newblock Predictive processing: a canonical cortical computation.
\newblock {\em Neuron}, 100(2):424--435, 2018.

\bibitem{rao1999predictive}
R.~P. Rao and D.~H. Ballard.
\newblock Predictive coding in the visual cortex: a functional interpretation
  of some extra-classical receptive-field effects.
\newblock {\em Nature neuroscience}, 2(1):79--87, 1999.

\bibitem{friston2010free}
K.~Friston.
\newblock The free-energy principle: a unified brain theory?
\newblock {\em Nature Reviews Neuroscience}, 11(2):127--138, 2010.

\bibitem{clark2015surfing}
A.~Clark.
\newblock {\em Surfing uncertainty: Prediction, action, and the embodied mind}.
\newblock Oxford University Press, 2015.

\bibitem{wacongne2012neuronal}
C.~Wacongne, J.-P. Changeux, and S.~Dehaene.
\newblock A neuronal model of predictive coding accounting for the mismatch
  negativity.
\newblock {\em Journal of Neuroscience}, 32(11):3665--3678, 2012.

\bibitem{bastos2012canonical}
A.~M. Bastos, W.~M. Usrey, R.~A. Adams, G.~R. Mangun, P.~Fries, and K.~J.
  Friston.
\newblock Canonical microcircuits for predictive coding.
\newblock {\em Neuron}, 76(4):695--711, 2012.

\bibitem{rao200216}
R.~P. Rao and T.~J. Sejnowski.
\newblock Predictive coding, cortical feedback, and spike-timing dependent
  plasticity.
\newblock {\em Probabilistic models of the brain}, page 297, 2002.

\bibitem{bogacz2017tutorial}
R.~Bogacz.
\newblock A tutorial on the free-energy framework for modelling perception and
  learning.
\newblock {\em Journal of mathematical psychology}, 76:198--211, 2017.

\bibitem{whittington2019theories}
J.~C. Whittington and R.~Bogacz.
\newblock Theories of error back-propagation in the brain.
\newblock {\em Trends in Cognitive Sciences}, 23(3):235--250, 2019.

\bibitem{hertag2020learning}
L.~Hert{\"a}g and H.~Sprekeler.
\newblock Learning prediction error neurons in a canonical interneuron circuit.
\newblock {\em Elife}, 9:e57541, 2020.

\bibitem{schulz2021generation}
A.~Schulz, C.~Miehl, M.~J. Berry~II, and J.~Gjorgjieva.
\newblock The generation of cortical novelty responses through inhibitory
  plasticity.
\newblock {\em Elife}, 10:e65309, 2021.

\bibitem{castillo2011long}
P.~E. Castillo, C.~Q. Chiu, and R.~C. Carroll.
\newblock Long-term plasticity at inhibitory synapses.
\newblock {\em Current opinion in neurobiology}, 21(2):328--338, 2011.

\bibitem{Vogels2011}
T.~P. Vogels, H.~Sprekeler, F.~Zenke, C.~Clopath, and W.~Gerstner.
\newblock {Inhibitory plasticity balances excitation and inhibition in sensory
  pathways and memory networks.}
\newblock {\em Science}, 334(6062):1569--73, dec 2011.

\bibitem{luz2012balancing}
Y.~Luz and M.~Shamir.
\newblock Balancing feed-forward excitation and inhibition via hebbian
  inhibitory synaptic plasticity.
\newblock {\em PLoS computational biology}, 8(1):e1002334, 2012.

\bibitem{Vogels2013}
T.~P. Vogels, R.~C. Froemke, N.~Doyon, M.~Gilson, J.~S. Haas, R.~Liu,
  A.~Maffei, P.~Miller, C.~J. Wierenga, M.~A. Woodin, F.~Zenke, and
  H.~Sprekeler.
\newblock {Inhibitory synaptic plasticity: spike timing-dependence and putative
  network function}.
\newblock {\em Frontiers in Neural Circuits}, 7(119), 2013.

\bibitem{Hennequin2017}
G.~Hennequin, E.~J. Agnes, and T.~P. Vogels.
\newblock {Inhibitory Plasticity: Balance, Control, and Codependence}.
\newblock {\em Annu. Rev. Neurosci.}, 40(1):557--579, 2017.

\bibitem{capogna2021ins}
M.~Capogna, P.~E. Castillo, and A.~Maffei.
\newblock The ins and outs of inhibitory synaptic plasticity: Neuron types,
  molecular mechanisms and functional roles.
\newblock {\em European Journal of Neuroscience}, 54(8):6882--6901, 2021.

\bibitem{baker2020nonlinear}
C.~Baker, V.~Zhu, and R.~Rosenbaum.
\newblock Nonlinear stimulus representations in neural circuits with
  approximate excitatory-inhibitory balance.
\newblock {\em PLoS computational biology}, 16(9):e1008192, 2020.

\bibitem{hertag2021prediction}
L.~Hert{\"a}g and C.~Clopath.
\newblock Prediction-error neurons in circuits with multiple neuron types:
  Formation, refinement and functional implications.
\newblock {\em bioRxiv}, 2021.

\bibitem{akil2021balanced}
A.~E. Akil, R.~Rosenbaum, and K.~Josi{\'c}.
\newblock Balanced networks under spike-time dependent plasticity.
\newblock {\em PLoS Computational Biology}, 17(5):e1008958, 2021.

\bibitem{brette2005adaptive}
R.~Brette and W.~Gerstner.
\newblock Adaptive exponential integrate-and-fire model as an effective
  description of neuronal activity.
\newblock {\em J Neurophysiol}, 94(5):3637--3642, 2005.

\bibitem{gerstner2014neuronal}
W.~Gerstner, W.~M. Kistler, R.~Naud, and L.~Paninski.
\newblock {\em Neuronal dynamics: From single neurons to networks and models of
  cognition}.
\newblock Cambridge University Press, 2014.

\bibitem{van2005course}
C.~van Vreeswijk and H.~Sompolinsky.
\newblock Methods and models in neurophysics course 9: Irregular activity in
  large networks of neurons.
\newblock {\em Les Houches}, 80:341--406, 2005.

\bibitem{vanVreeswijk:1996us}
C.~van Vreeswijk and H.~Sompolinsky.
\newblock {Chaos in neuronal networks with balanced excitatory and inhibitory
  activity}.
\newblock {\em Science}, 274(5293):1724--1726, 1996.

\bibitem{Amit:1997uj}
D.~Amit and N.~Brunel.
\newblock {Model of global spontaneous activity and local structured activity
  during delay periods in the cerebral cortex}.
\newblock {\em Cereb Cortex}, 7(3):237--252, 1997.

\bibitem{vanVreeswijk:1998uz}
C.~van Vreeswijk and H.~Sompolinsky.
\newblock {Chaotic balanced state in a model of cortical circuits}.
\newblock {\em Neural Comput}, 10(6):1321--1371, 1998.

\bibitem{Brunel:1999ua}
N.~Brunel and V.~Hakim.
\newblock {Fast global oscillations in networks of integrate-and-fire neurons
  with low firing rates}.
\newblock {\em Neural Comput}, 11(7):1621--1671, 1999.

\bibitem{Brunel:2000th}
N.~Brunel.
\newblock Dynamics of sparsely connected networks of excitatory and inhibitory
  spiking neurons.
\newblock {\em J Comput Neurosci}, 8(3):183--208, 2000.

\bibitem{Renart2010}
A.~Renart, J.~de~La~Rocha, P.~Bartho, L.~Hollender, N.~Parga, A.~Reyes, and
  K.~Harris.
\newblock {The Asynchronous State in Cortical Circuits}.
\newblock {\em Science}, 327(5965):587--590, 2010.

\bibitem{pyle2016highly}
R.~Pyle and R.~Rosenbaum.
\newblock Highly connected neurons spike less frequently in balanced networks.
\newblock {\em Phys Rev E}, 93(4):040302(R), 2016.

\bibitem{pyle2017spatiotemporal}
R.~Pyle and R.~Rosenbaum.
\newblock Spatiotemporal dynamics and reliable computations in recurrent
  spiking neural networks.
\newblock {\em Physical Rev Lett}, 118(1):018103, 2017.

\bibitem{ebsch2018imbalanced}
C.~Ebsch and R.~Rosenbaum.
\newblock Imbalanced amplification: A mechanism of amplification and
  suppression from local imbalance of excitation and inhibition in cortical
  circuits.
\newblock {\em PLoS Comp Bio}, 14(3):e1006048, 2018.

\bibitem{baker2019correlated}
C.~Baker, C.~Ebsch, I.~Lampl, and R.~Rosenbaum.
\newblock Correlated states in balanced neuronal networks.
\newblock {\em Phys Rev E}, 99(5):052414, 2019.

\bibitem{saxena2019towards}
S.~Saxena and J.~P. Cunningham.
\newblock Towards the neural population doctrine.
\newblock {\em Current opinion in neurobiology}, 55:103--111, 2019.

\bibitem{jordan2020opposing}
R.~Jordan and G.~B. Keller.
\newblock Opposing influence of top-down and bottom-up input on excitatory
  layer 2/3 neurons in mouse primary visual cortex.
\newblock {\em Neuron}, 108(6):1194--1206, 2020.

\bibitem{Rosenbaum2014}
R.~Rosenbaum and B.~Doiron.
\newblock {Balanced networks of spiking neurons with spatially dependent
  recurrent connections}.
\newblock {\em Phys Rev X}, 4(2):021039, 2014.

\bibitem{lim:2014}
S.~Lim and M.~S. Goldman.
\newblock Balanced cortical microcircuitry for spatial working memory based on
  corrective feedback control.
\newblock {\em J Neurosci.}, 34(20):6790--6806, 2014.

\bibitem{landau2016impact}
I.~D. Landau, R.~Egger, V.~J. Dercksen, M.~Oberlaender, and H.~Sompolinsky.
\newblock The impact of structural heterogeneity on excitation-inhibition
  balance in cortical networks.
\newblock {\em Neuron}, 92(5):1106--1121, 2016.

\bibitem{ahmadian2021dynamical}
Y.~Ahmadian and K.~D. Miller.
\newblock What is the dynamical regime of cerebral cortex?
\newblock {\em Neuron}, 109(21):3373--3391, 2021.

\bibitem{turrigiano2011too}
G.~Turrigiano.
\newblock Too many cooks? intrinsic and synaptic homeostatic mechanisms in
  cortical circuit refinement.
\newblock {\em Annual review of neuroscience}, 34:89--103, 2011.

\end{thebibliography}
\end{document}